\documentclass[onecolumn]{emulateapj}
\usepackage{color}
\usepackage{epsfig}
\def\lsim{\raise0.3ex\hbox{$<$}\kern-0.75em{\lower0.65ex\hbox{$\sim$}}}
\def\gsim{\raise0.3ex\hbox{$>$}\kern-0.75em{\lower0.65ex\hbox{$\sim$}}}

\begin{document}

\title{Wide-Field $^{12}$CO ($J=2-1$) and $^{13}$CO ($J=2-1$) Observations toward the Aquila Rift and 
Serpens Molecular Cloud Complexes.
I. Molecular Clouds and Their Physical Properties}
\author{Fumitaka Nakamura\altaffilmark{1,2,3}, 
Kazuhito Dobashi\altaffilmark{4}, Tomomi Shimoikura\altaffilmark{4},
Tomohiro Tanaka\altaffilmark{5}, Toshikazu Onishi\altaffilmark{5}}
\altaffiltext{1}{National Astronomical Observatory of Japan, 2-21-1 Osawa, Mitaka, Tokyo 181-8588}
\altaffiltext{2}{The Graduate University for Advanced Studies (SOKENDAI), 2-21-1 Osawa, Mitaka, Tokyo 181-0015, Japan}
\altaffiltext{3}{Nobeyama Radio Observatory, National Astronomical Observatory of Japan
462-2 Nobeyama, Minamimaki, Minamisaku, Nagano 384-1305}
 \altaffiltext{4}{Department of Astronomy and Earth Sciences, 
Tokyo Gakugei University, 4-1-1 Nukuikitamachi, Koganei, Tokyo 184-8501}
 \altaffiltext{5}{Department of Physical Science, Graduate School of
   Science, Osaka Prefecture University, 1-1 Gakuen-cho, Naka-ku, Sakai,
   Osaka}

\begin{abstract}
We present results of wide-field $^{12}$CO ($J = 2 - 1$) and
$^{13}$CO ($J = 2 - 1$) observations toward the Aquila Rift and Serpens
molecular cloud complexes (25$^\circ < l < 33^\circ$ and $1^\circ < b < 6^\circ$) 
at an angular resolution of 3$'$.4 ($\approx$ 0.25 pc) and at a velocity resolution 
of 0.079 km s$^{-1}$ with the velocity coverage of $-5$ km s$^{-1} < V_{\rm LSR} <$ 35
km s$^{-1}$. 
We found that the $^{13}$CO emission better traces the structures seen in the extinction map and
derived the $X_{\rm ^{13}CO}$-factor of this region. 
Applying \texttt{SCIMES} to the $^{13}$CO data cube, we identified 61 clouds and derived 
their masses, radii, and line widths. 
The line-width-radius relation of the identified clouds  basically follows 
those of nearby molecular clouds.
Majority of the identified clouds are close to virial equilibrium although the dispersion is large.
By inspecting the $^{12}$CO channel maps by eye, we found several arcs which are spatially extended to 0.2 $-$ 3 degree in length. 
In the longitude-velocity diagrams of $^{12}$CO, we also found the two spatially-extended components which appear to converge toward Serpens South and W40 region.  The existence of two components with different velocities and arcs suggests that large-scale expanding bubbles and/or flows play a role 
in the formation and evolution of the Serpens South and W40 cloud.   
\end{abstract}
\keywords{ISM: clouds --- ISM: kinematics and dynamics --- 
ISM: molecules --- ISM: structure --- stars: formation}

\section{Introduction}
\label{sec:intro}

Large-scale molecular line observations are important to
unveil how molecular clouds have been formed and evolved
because in interstellar space, large-scale dynamical events such as
bubbles and supersonic turbulent flows often influence the structure
and kinematics of molecular gas where star formation happens
\citep[e.g.,][]{dame85,dame87}. 
In the present paper, we investigate the cloud structure 
and kinematics of the Aquila Rift and Serpens molecular cloud complex, 
on the basis of large-scale multi-CO line observations.

Aquila Rift is located in the first Galactic
quadrant, spanning from 20$^\circ$ to 40$^\circ$ in Galactic
longitude and $-1^{\circ}$ to 10$^\circ$ in Galactic latitude.
It appears in the optical image as a dark lane that divides 
the bright band of the Milky Way longitudinally \citep{prato08}.
The total molecular gas mass is estimated to be about a few 
$\times 10^5 M_\odot$ on the basis of the CO ($J=1-0$) observations with an 
angular resolution of about 7$'$.5 \citep{dame85,dame87}.
\citet{dobashi05} identified a number of dark clouds in this region
using the Digitized Sky Survey visual extinction data.
These previous studies indicate that the region has  complex
density and velocity structure, suggesting that the dynamical 
interaction and events might be ongoing and/or have happened in this region
\citep{prato08}. 
In fact, \citet{frisch98} pointed out that several nearby superbubble shells appear to 
converge toward the Aquila Rift region.
In the southern part of Aquila Rift, \citet{kawamura99} suggested
that the virial mass of the molecular cloud complex is significantly
larger than the molecular gas mass and star formation is less active.

However, the molecular gas distribution and star formation activity in
the northern part above the Galactic plane have not been extensively explored so far.
Here, we focus on the western part of Aquila Rift, which stretches from 
$25^\circ$ to $33^\circ$ in Galactic longitude and 
$1^\circ$ to $6^\circ$ in Galactic latitude.
The western part of Aquila Rift contains several active star-forming regions
such as Serpens Main, Serpens South, W40, and MWC 297.
In a well-studied nearby cluster-forming region, Serpens main star-forming
region \citep{eiroa08}, several young protostars are blowing out of powerful
collimated outflows.
Recently, the {\it Spitzer} observations have discovered 
an extremely-young embedded cluster of low-mass protostars, Serpens
South, which contains a large number of Class 0/I objects \citep{gutermuth08,andre10,tanaka13,konyves15}.
In fact, \citet{nakamura11} detected a number of CO outflows in the central part
of Serpens South \citep[see also][]{plunkett15}.
In more evolved star-forming region, W40 H\,{\sc ii} region, the expanding structure affects the star formation
in this region \citep{shimoikura15}.
Recent studies also suggest that cloud-cloud collision may have triggered 
star formation in Serpens Main region \citep{duarte11} and
Serpens South region \citep{nakamura14}.
MWC 297 is an embedded young massive B1.5Ve star, one of the closest massive stars
\citep{drew97}.  A molecular outflow is detected toward the MWC 297 region, 
implying that active star formation is ongoing.
In spite of the effort of these previous studies, it remains uncertain  why star formation in the western part is much more active than the southern part.

In addition, the distance to the Aquila Rift and Serpens cloud complex is somewhat controversial.
For the Serpens Main star-forming region, the distance of $\sim$ 260 pc has been used \citep{eiroa08}. 
This distance determination is based mainly on measurements of the extinction suffered by stars 
in the direction of Serpens. 
However, recent VLBA measurements of young stellar objects, EC95a and EC95b, 
suggest a larger distance of 415 pc \citep{dzib10}.
For the W40 cloud, the distance is estimated to be $300-900$ pc 
and has not yet been determined to a satisfactory precision \citep{rodney08}.
The distance to MWC 297 is estimated to be 250 pc \citep{drew97} or 450 pc \citep{hillenbrand92}
on the basis of measurements of the extinction of stars, and the uncertainty is a similar degree
to other regions in the Aquila Rift and Serpens cloud complexes.
Serpens South has a Local Standard of Rest (LSR) velocity similar to Serpens Main \citep{gutermuth08}.
This fact suggests that Serpens South may have a distance similar to Serpens Main.  
However, no accurate distance measurements have been done so far.
The uncertainty of the distance makes it difficult to clarify the cloud dynamical states in the
Aquila Rift and Serpens cloud complexes.
In the present paper, we compare the cloud physical quantities  for both 260 pc and 415 pc.

In the present paper, as a first step toward a better understanding of star formation activity
in the Aquila Rift and Serpens molecular cloud complexes,
we investigate the large-scale molecular cloud structure of the region 
with an angular resolution of about 3$'$, on the basis
of wide field $^{12}$CO ($J=2-1$) and  $^{13}$CO ($J=2-1$) observations.
In Section \ref{sec:obs}, we present the detail of our observations. 
In Section \ref{sec:results}, we describe the large-scale CO structure in this region. 
We find several arcs that may have been formed by the large-scale flows. 
We also derive the $X_{\rm CO}$ factor of this region by comparing CO velocity-integrated 
intensity map and the 2MASS extinction map.
In Section \ref{sec:clouds}, we apply \texttt{SCIMES}  \citep{colombo15} to the CO data cube, 
and identify the clouds.  In the present paper, we call structures identified by \texttt{SCIMES} as "clouds".
Then, we attempt  to assess their dynamical states. 
Finally, we briefly summarize the main results in Section \ref{sec:summary}.

\section{Observations and Data}
\label{sec:obs}

\subsection{1.85-m Observations}
We carried out $^{12}$CO ($J=2-1$), $^{13}$CO ($J=2-1$), and C$^{18}$O ($J=2-1$)
 mapping observations toward the Aquila Rift and Serpens molecular cloud complexes,
during the periods from 2012 February to 2012 March and from 2012 December to 2013 March.
The observations were done in the on-the-fly (OTF) mapping mode with a 2SB SIS mixer receiver on the 1.85 m
 radio telescope of Osaka Prefecture University. The telescope is installed at the Nobeyama Radio Observatory (NRO).
The image rejection ratio (IRR) was measured to be 10 dB or higher during the observations.
 At 230 GHz band, the telescope has a beam size of $2'.7$ (HPBW) 
and a main beam efficiency of $\eta$ = 0.6.
 At the back end, we used a digital Fourier transform spectrometer (DFS) 
 with 16384 channels that covers the 1 GHz bandwidth,
 which allows us to obtain the three molecular lines,  
 $^{12}$CO ($J=2-1$), $^{13}$CO ($J=2-1$), and C$^{18}$O ($J=2-1$), simultaneously.
 In the present paper, we focus on the $^{12}$CO ($J=2-1$) and $^{13}$CO ($J=2-1$) data.
 The frequency resolution was set to 61 kHz, which corresponds to the velocity resolution of 
$\sim$ 0.08 km s$^{-1}$ at 220 GHz.
 During the observations, the system noise temperatures were about
 200 $-$ 400 K in a single sideband.
Further description of the telescope is given by \citet{onishi13}.

The observed area was a rectangle area with 40 square degrees whose
coordinates of the bottom-left-corner (BLC) and top-right-corner
(TRC) are  $(l, b) \simeq$ ($33^\circ$, $1^\circ$) and ($25^\circ$,
$6^\circ$), respectively.
The area was divided into 40 boxes of $1^\circ \times 1^\circ $, each of
which was scanned a few times in both Galactic longitude and lattitude 
with a row spacing of $1'$ at a scan speed of $1$ sec per row.
The standard chopper wheel method was used to correct
the output signals for the atmospheric attenuation and to
convert them into the antenna temperatures ($T_A^*$). 
Then, we corrected the antenna temperatures for the main beam efficiency to obtain
the brightness temperatures $T_{\rm mb}$ (= $T_A^*/\eta$).
We applied a convolution technique with a Gaussian function to calculate
the intensity onto a regular grid with a spacing of 1$'$.
The resultant effective angular resolution was $3'.4$.
The rms noise levels of the final $^{12}$CO and $^{13}$CO maps vary from box to box, and
are summarized in Appendix A (Tables A1 and A2).
The average noise levels are given in the last column of Table 1.

\subsection{2MASS Extinction Data}

\begin{figure}
\centering
\includegraphics[width=100mm]{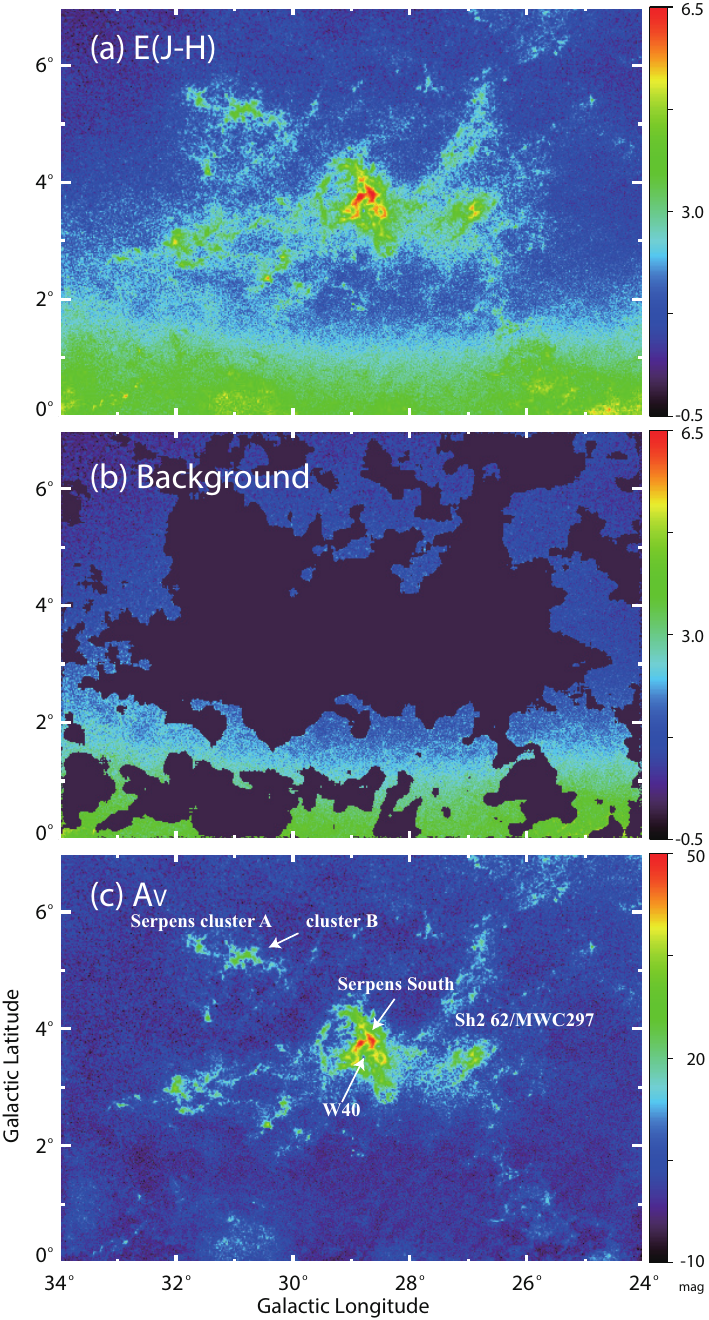}
\caption{(a) Color excess map of $E(J-H)$ of the Aquila Rift and Serpens
molecular cloud complex generated by \citet{dobashi11}
and \citet{dobashi13}. The fits data were downloaded from their website
(http://darkclouds.u-gakugei.ac.jp/2MASS/download.html).
(b) Same as (a), but the map is masked for small-scale structures 
to determine the background.
(c) Extinction map of $A_V$ derived from the background-subtracted $E(J-H)$ map.
The names of the regions discussed in the text are designated in the panel.
}  
\label{fig:2mass}
\end{figure}

To compare with the molecular line data, we downloaded the near-infrared (NIR) 
color excess map of $E(J-H)$ shown in Figure \ref{fig:2mass}(a) from http://darkclouds.u-gakugei.ac.jp/2MASS/download.html. The map was
originally generated by \citet{dobashi11} and \citet{dobashi13} 
utilizing the 2MASS point source catalog, and sampled at the same $1'$ 
grid along the galactic coordinates as the CO data. 
The angular resolution, however, varies in the range of $0.8' - 11'$, 
because the map was drawn using the ''adaptive grid" technique 
to achieve a constant noise level. See \citet{dobashi11} and \citet{dobashi13} for details.
In Section \ref{sec:av}, using the $E(J-H)$ color excess map, we construct the visual extinction map 
of the Aquila Rift and Serpens cloud complexes, which is presented in Figure \ref{fig:2mass}(c).
Some of the active star-forming regions are designated in Figure  \ref{fig:2mass}(c).

\section{Results}
\label{sec:results}

\subsection{Global Molecular Gas Distribution}
\label{sec:global}

\begin{figure}
\plotone{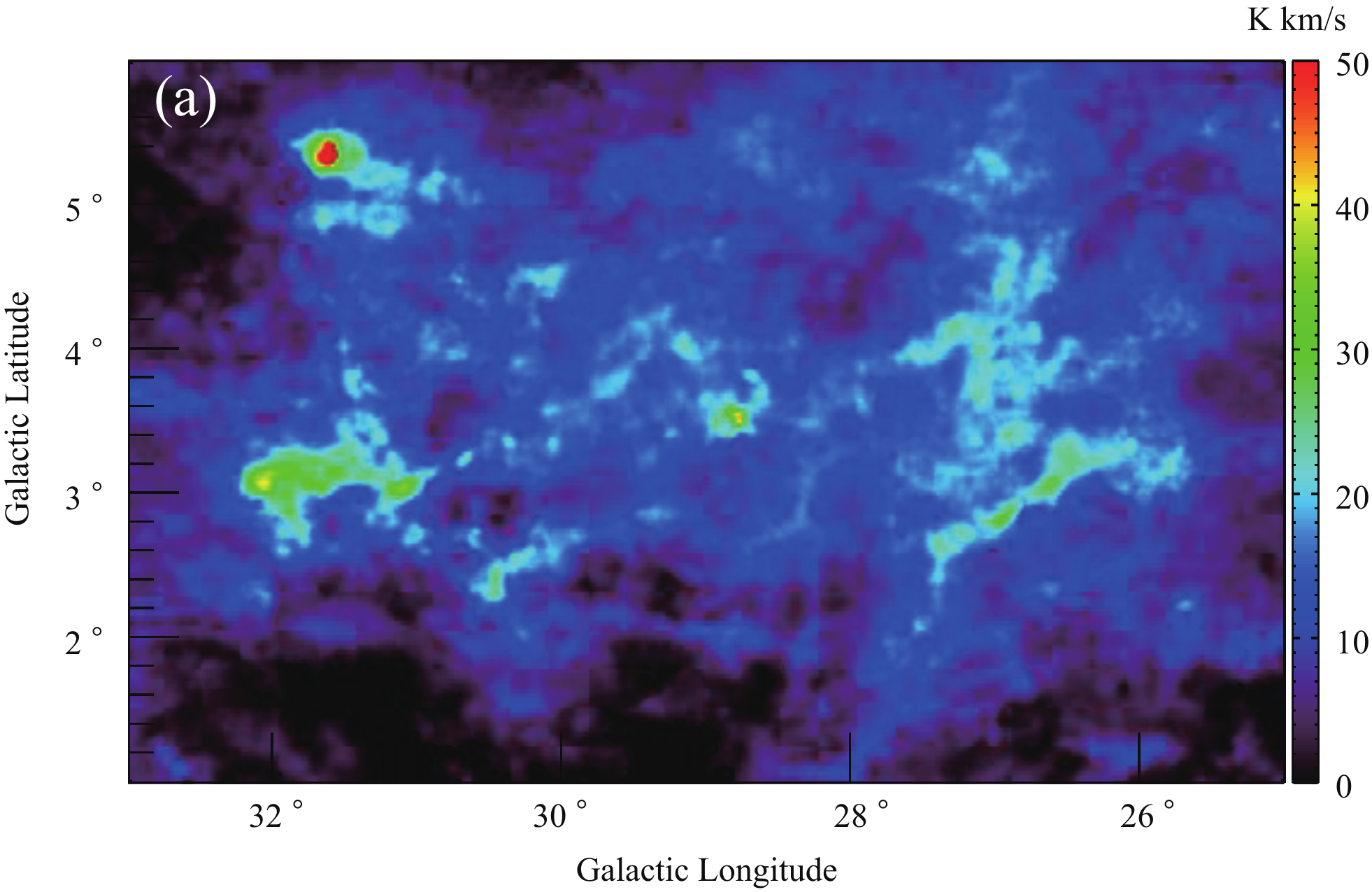}
\plotone{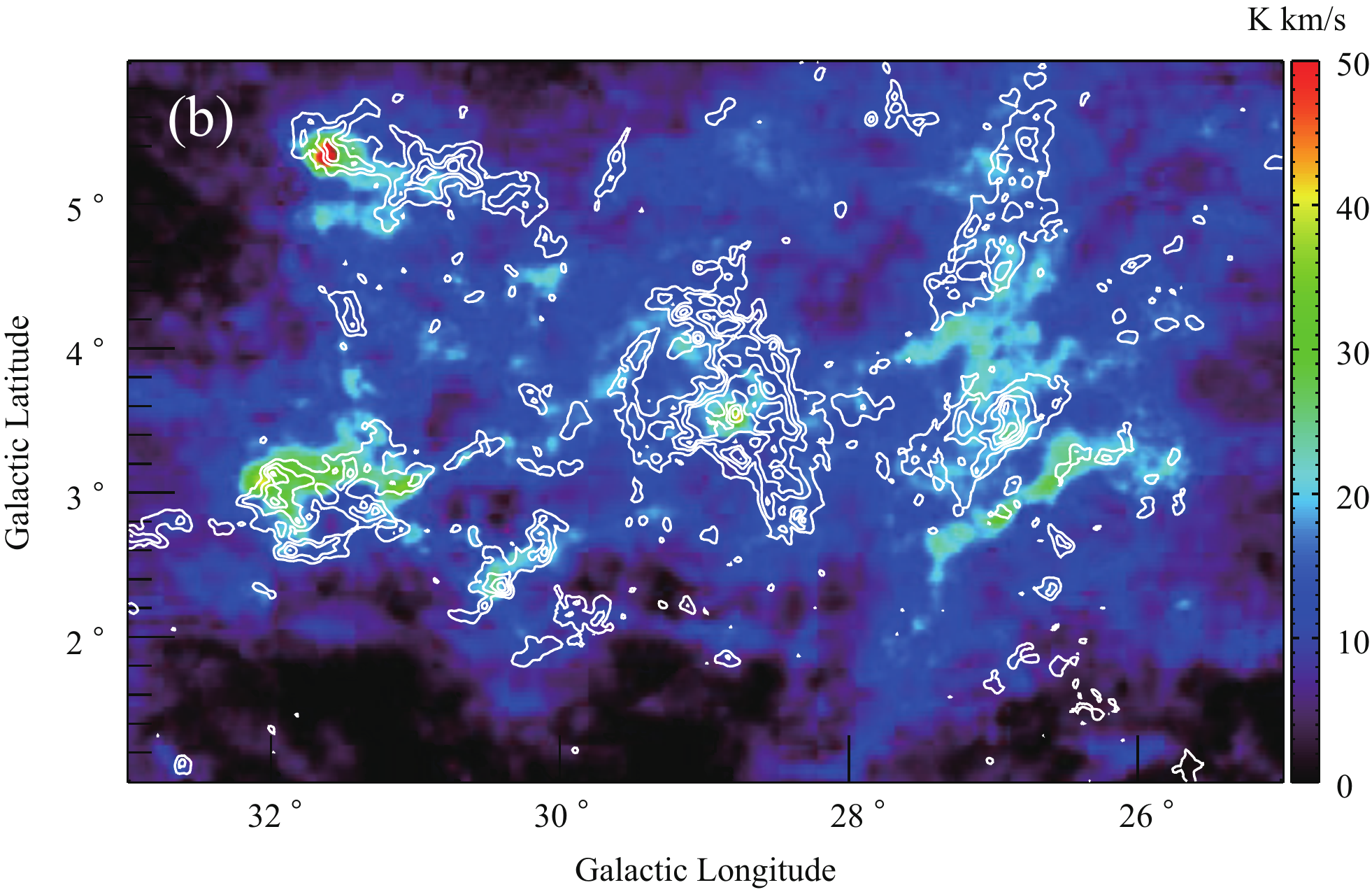}
\caption{
(a) $^{12}$CO ($J=2-1$) total integrated intensity map in the velocity 
range from  $V_{\rm LSR} = 0 $ km s$^{-1}$ to
20 km s$^{-1}$ toward the Aquila Rift and Serpens molecular cloud
 complexes.
(b) Same as panel (a) but with contours of the 2MASS $A_V$ map.
The contours start at 2.0 mag with intervals of 3.0 mag.
}  
\label{fig:12CO integ map}
\end{figure}

\begin{figure}
\plotone{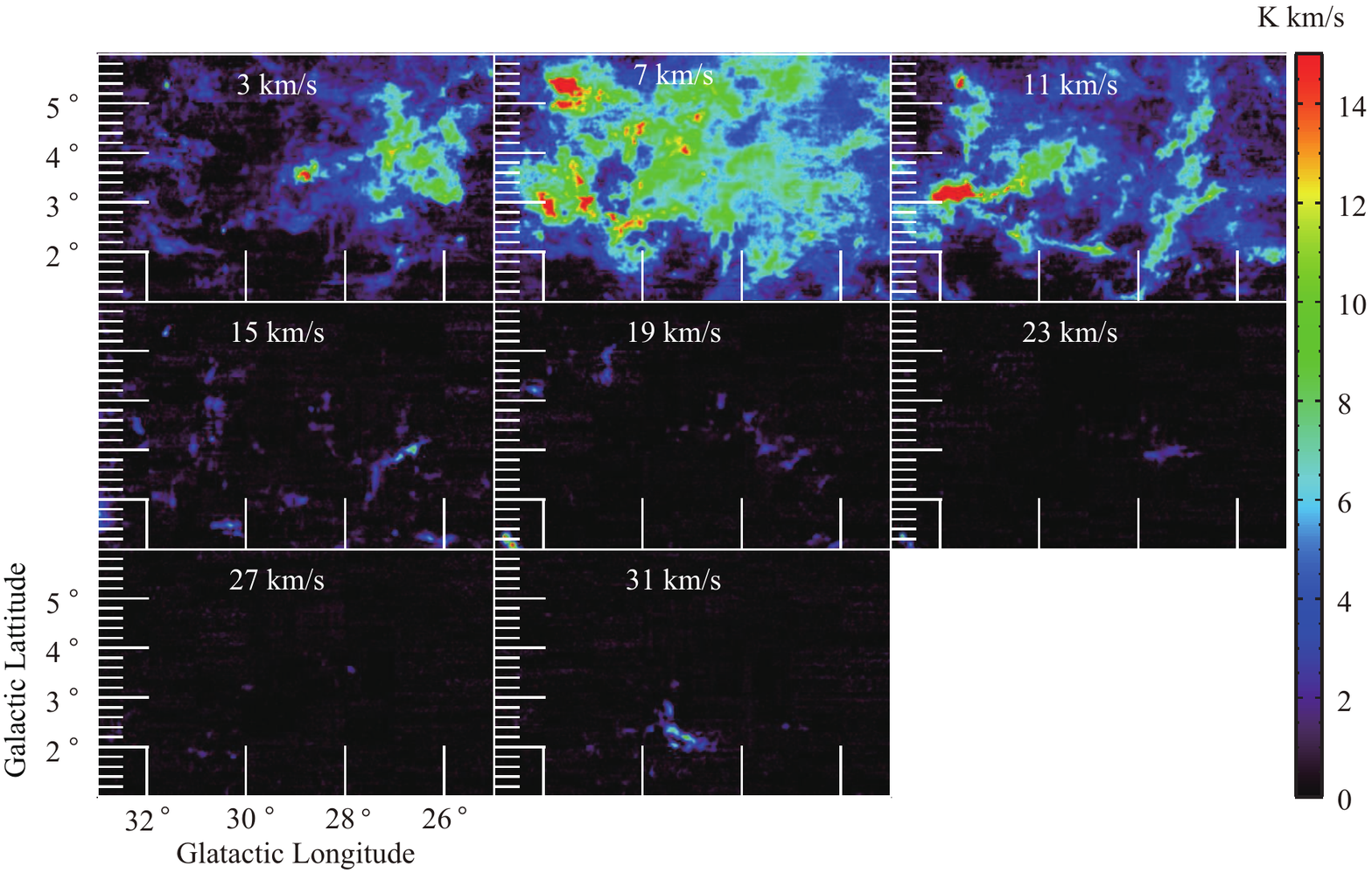}
\caption{
$^{12}$CO ($J=2-1$) velocity channel maps
toward the Aquila Rift and Serpens molecular cloud complexes.
(Animations of the channel maps are available in the online journal.)}  
\label{fig:12CO channel map}
\end{figure}

\begin{figure}
\plotone{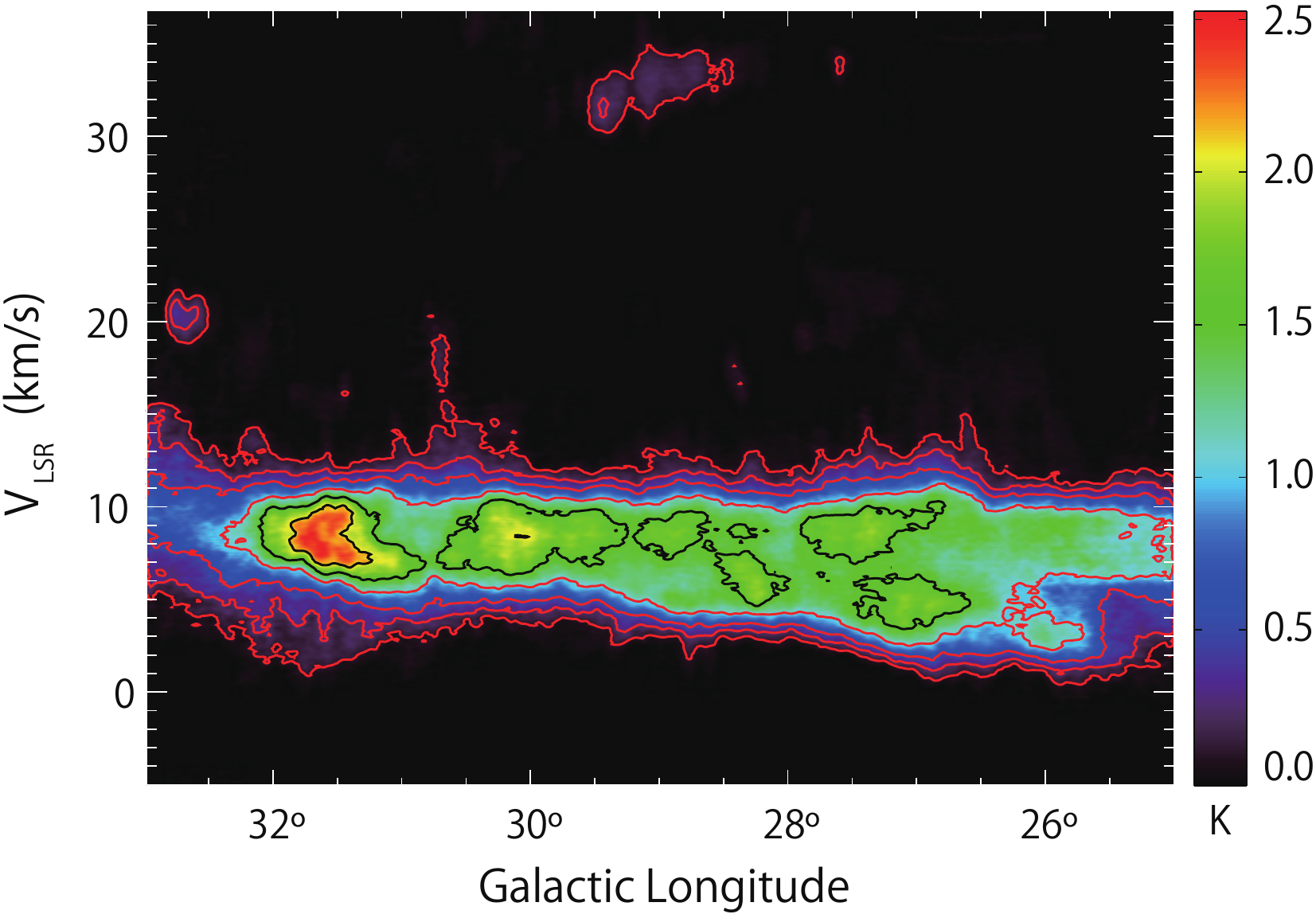}
\caption{
$^{12}$CO ($J=2-1$) longitude-velocity diagram integrated over the entire
 Galactic latitude range observed ($1^\circ < b < 6^\circ$).
 The contours are drawn at 0.1, 0.25, 0.5, 1.0, 1.5, 2.0, and 2.5 K.
}
\label{fig:PV map}
\end{figure}

\begin{figure}
\plotone{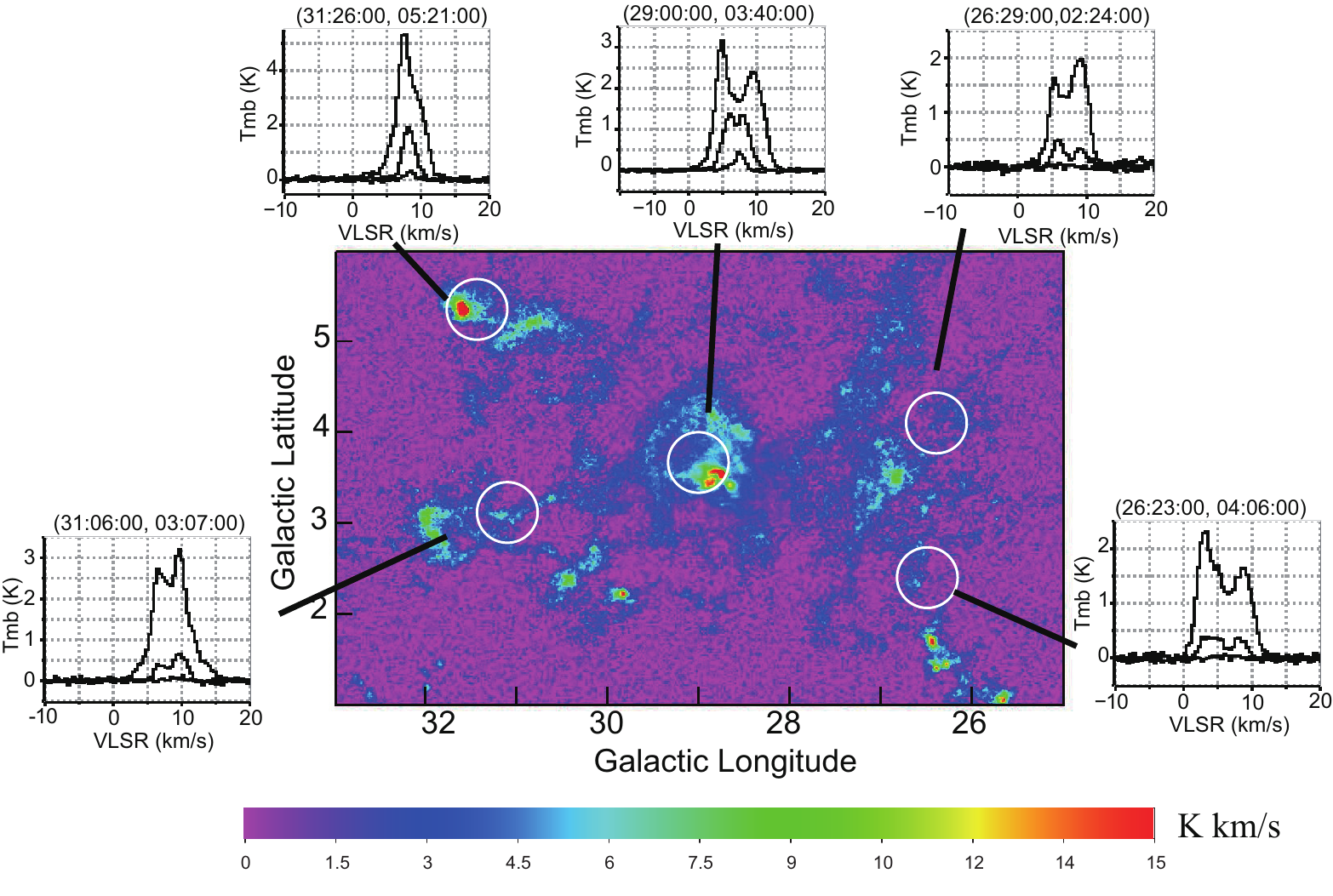}
\caption{CO line spectra at several positions.
Three CO line spectra are averaged in the circles with radius of 20$''$.
In each panel, the strongest, second strongest, and weakest lines are 
$^{12}$CO ($J=2-1$), $^{13}$CO ($J=2-1$), and C$^{18}$O ($J=2-1$), respectively.
The image is the $^{13}$CO ($J=2-1$) velocity integrated intensity map.
}
\label{fig:spectra}
\end{figure}

Figure \ref{fig:12CO integ map} shows the $^{12}$CO ($J=2-1$) velocity integrated
intensity map
toward the Aquila Rift and Serpens molecular cloud complexes.
For comparison, the $^{12}$CO ($J=2-1$) channel maps and the 
longitude-velocity diagram averaged over the Galactic latitude 
range of $1^\circ \le b \le 6^\circ$ are shown in 
Figures \ref{fig:12CO channel map} and  \ref{fig:PV map}, respectively.
$^{12}$CO, $^{13}$CO, and C$^{18}$O line spectra at several positions 
are also presented in Figure \ref{fig:spectra}.
The CO emission is extended almost over the entire mapped area.
The CO integrated intensity takes its maximum 
at the position of 
$(l, b) \simeq (31^\circ 35', \ 5^\circ 20')$, which 
is in the molecular cloud associated with the Serpens cluster A.
The $^{12}$CO emission associated with the Serpens cluster B is not so prominent.
Other CO peaks are located at
$(l, b) \simeq (28^\circ 45', \ 3^\circ 32.5')$, which corresponds to
the W40 H\,{\sc ii} region, and at the position of
$(l, b) \simeq (29^\circ 50', \ 2^\circ 13')$.
Most CO emission comes from the velocity range of 0 $-$ 20 km s$^{-1}$,
which is presumably associated with the Gould Belt, an expanding ring
of stars in the local spiral arm.

A prominent feature of the longitude-velocity and channel maps is the existence of two spatially-extended components 
with different velocities, $V_{\rm LSR} \sim 5 $ km s$^{-1}$ and 8 km s$^{-1}$ in the part
of Galactic longitude smaller than 29$^\circ$. 
The former component is recognized in the first panel ($V_{\rm LSR} = 3$ km s$^{-1}$) of the velocity channel map 
in Figure \ref{fig:12CO channel map}. The latter is more spatially-extended in the entire observed area.
In the $^{12}$CO longitude-velocity diagram (Figure \ref{fig:PV map}), 
the two components appear to converge at around the position of the Serpens South and W40 cloud,
which might suggest the two components interact with each other, triggering the active star formation
in Serrpens South and W40.
However, there is also a possibility that the two components are simply overlapped along the line-of sight.

\begin{figure}
\plotone{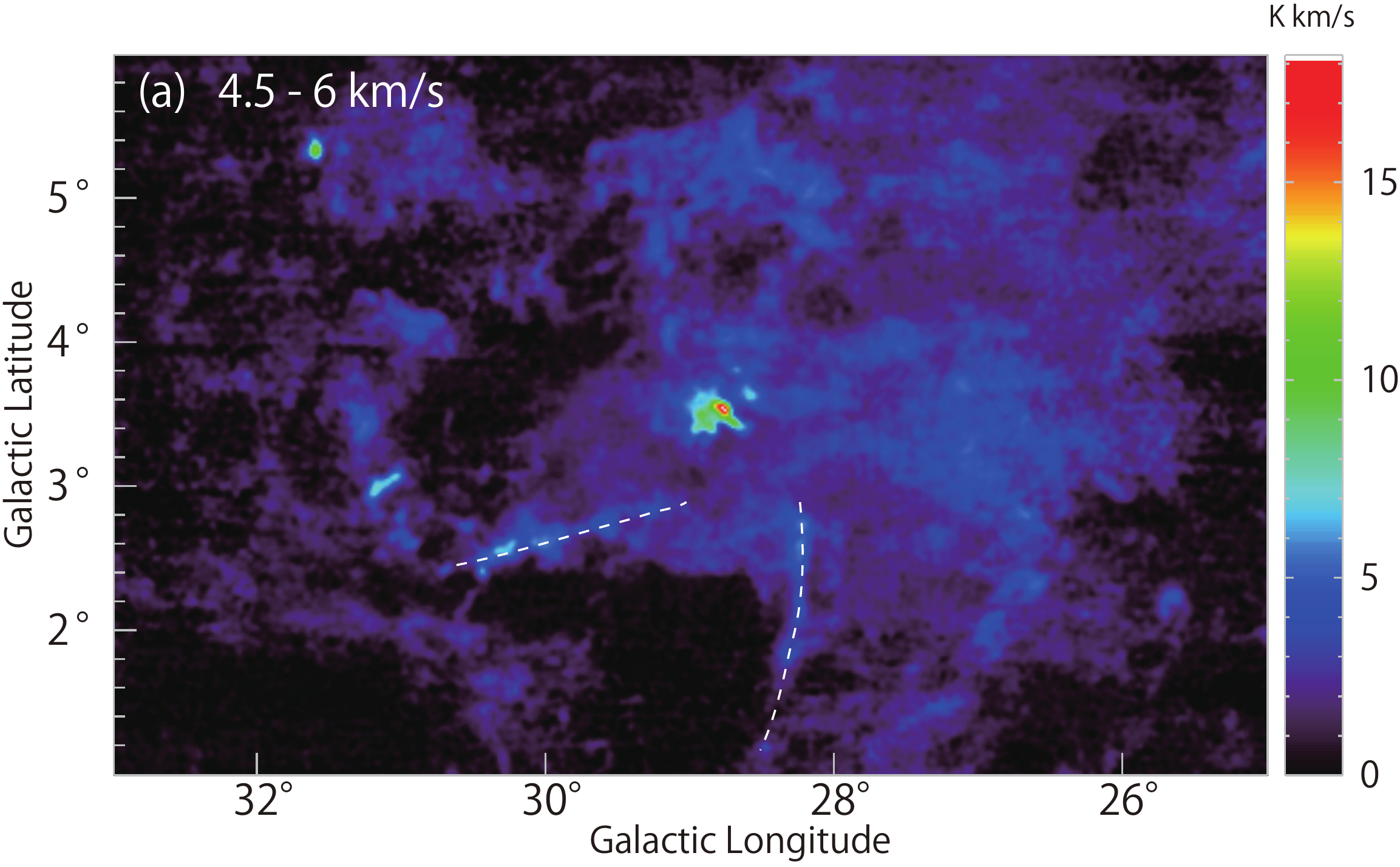}
\plotone{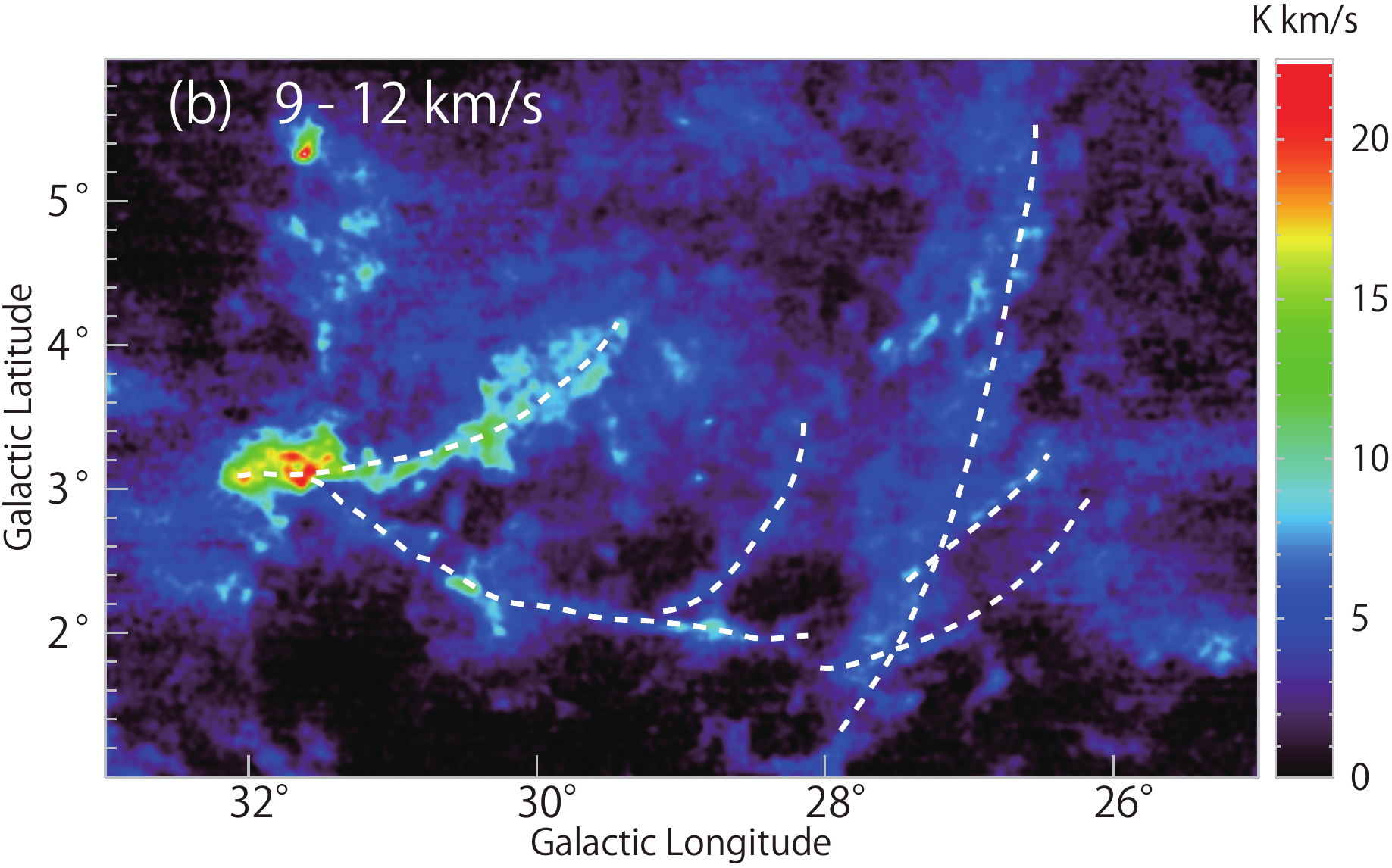}
\caption{
Total intensity maps of the $^{12}$CO ($J=2-1$) emission integrated 
over (a) 4.5 $< V_{\rm LSR} < $6.0 km s$^{-1}$ and 
(b) $9.0 < V_{\rm LSR} < 12.0$ km s$^{-1}$. 
}  
\label{fig:12CO channel map2}
\end{figure}

Another interesting characteristic is that in the channel maps,  some velocity-coherent, large-scale
linear structures can be recognized. 
For example, in both 1$-$4 km s$^{-1}$ (the first panel of Figure \ref{fig:12CO channel map}) 
and 9$-$13 km s$^{-1}$ (the third panel of Figure \ref{fig:12CO channel map}) panels, 
a couple of large arcs with similar morphologies are recognized.
To emphasize these structures, 
we show in Figures \ref{fig:12CO channel map2}a and \ref{fig:12CO channel map2}b, 
the $^{12}$CO intensity maps integrated over the velocity intervals 
of $4.5-6$ km s$^{-1}$ and $6-9$ km s$^{-1}$, respectively, where
the velocity-coherent linear structures are indicated with dashed curves.
Assuming a distance of 260 pc, these linear structures are found to be 
very large as about 6 $-$ 20 pc in length.
Hereafter, we call these linear structures arcs. 
Since ISM is highly turbulent, such large-scale structures 
should be created by large-scale dynamical events such as large-scale turbulent  
flows and/or supernova shocks.
These arcs might be created by the superbubbles converging toward 
the observed area \citep{frisch98}.

\begin{figure}
\plotone{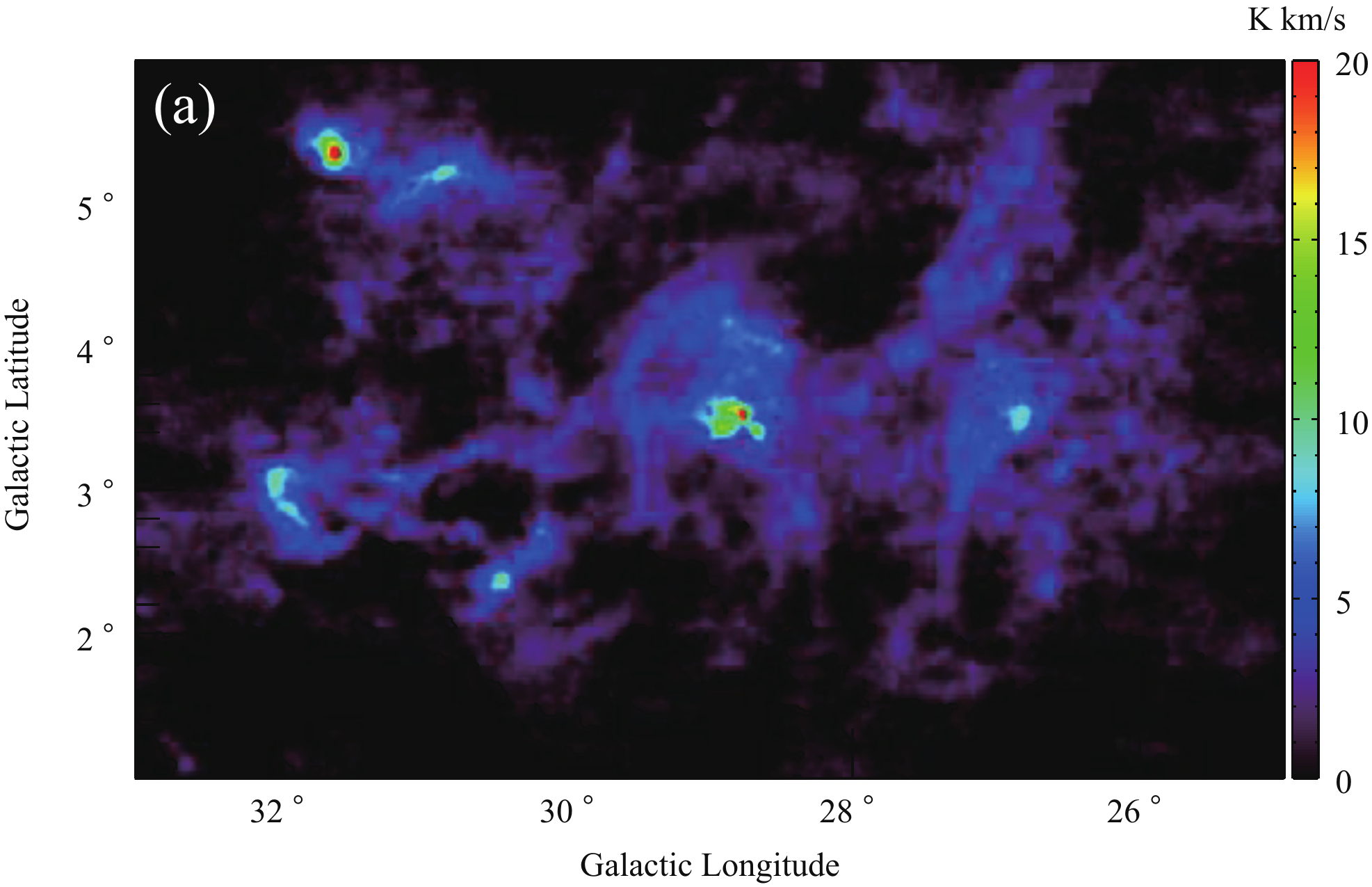}
\plotone{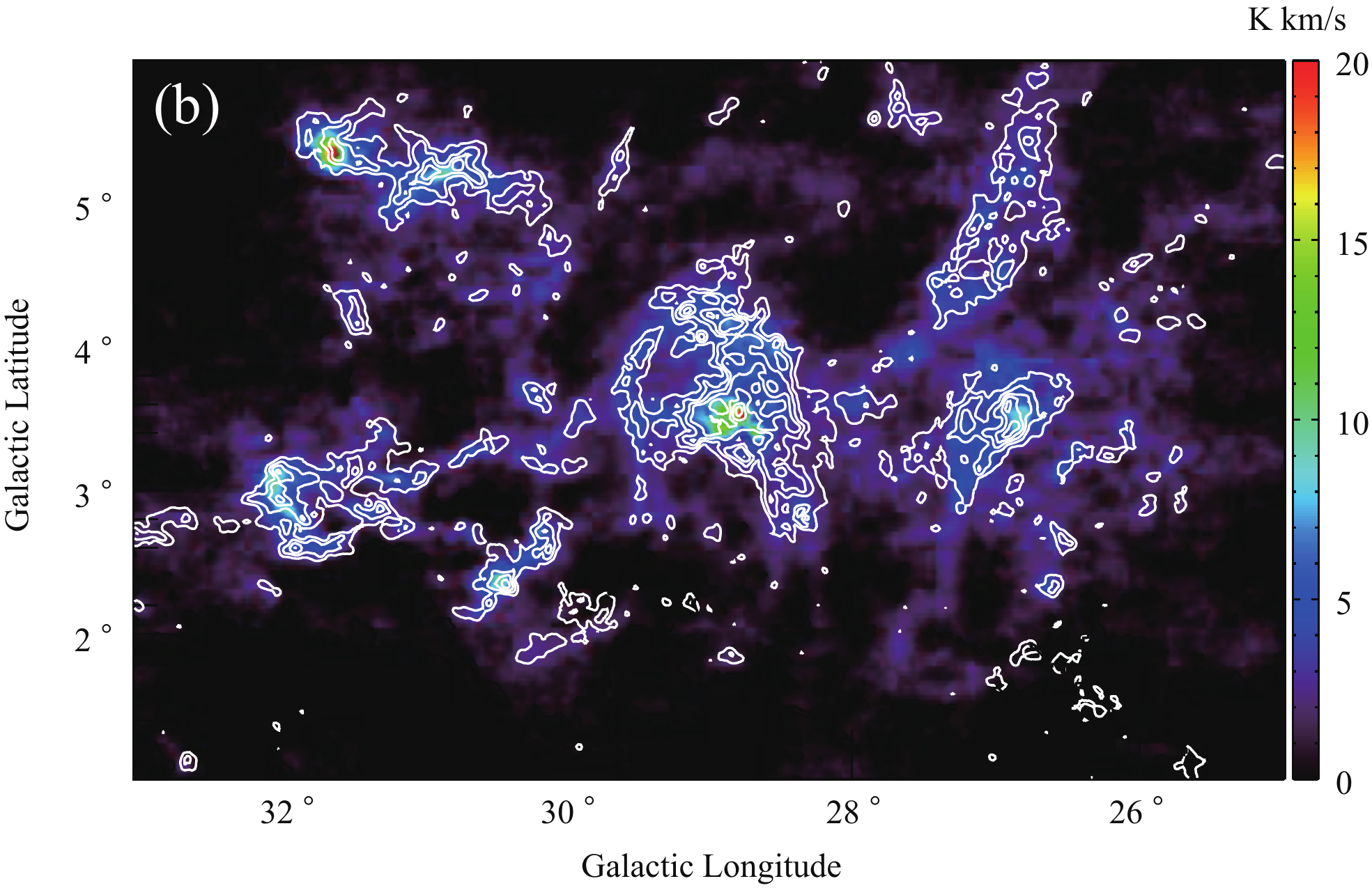}
\caption{
(a) $^{13}$CO ($J=2-1$) total integrated intensity map in the velocity range from 
 $V_{\rm LSR} = 0$  km s$^{-1}$ to 20 km s$^{-1}$ toward the Aquila Rift 
and Serpens molecular cloud complexes.
(b) Same as panel (a) but contours are the same as those of Figure 2b.
}  
\label{fig:13CO integ map}
\end{figure}

\begin{figure}
\plotone{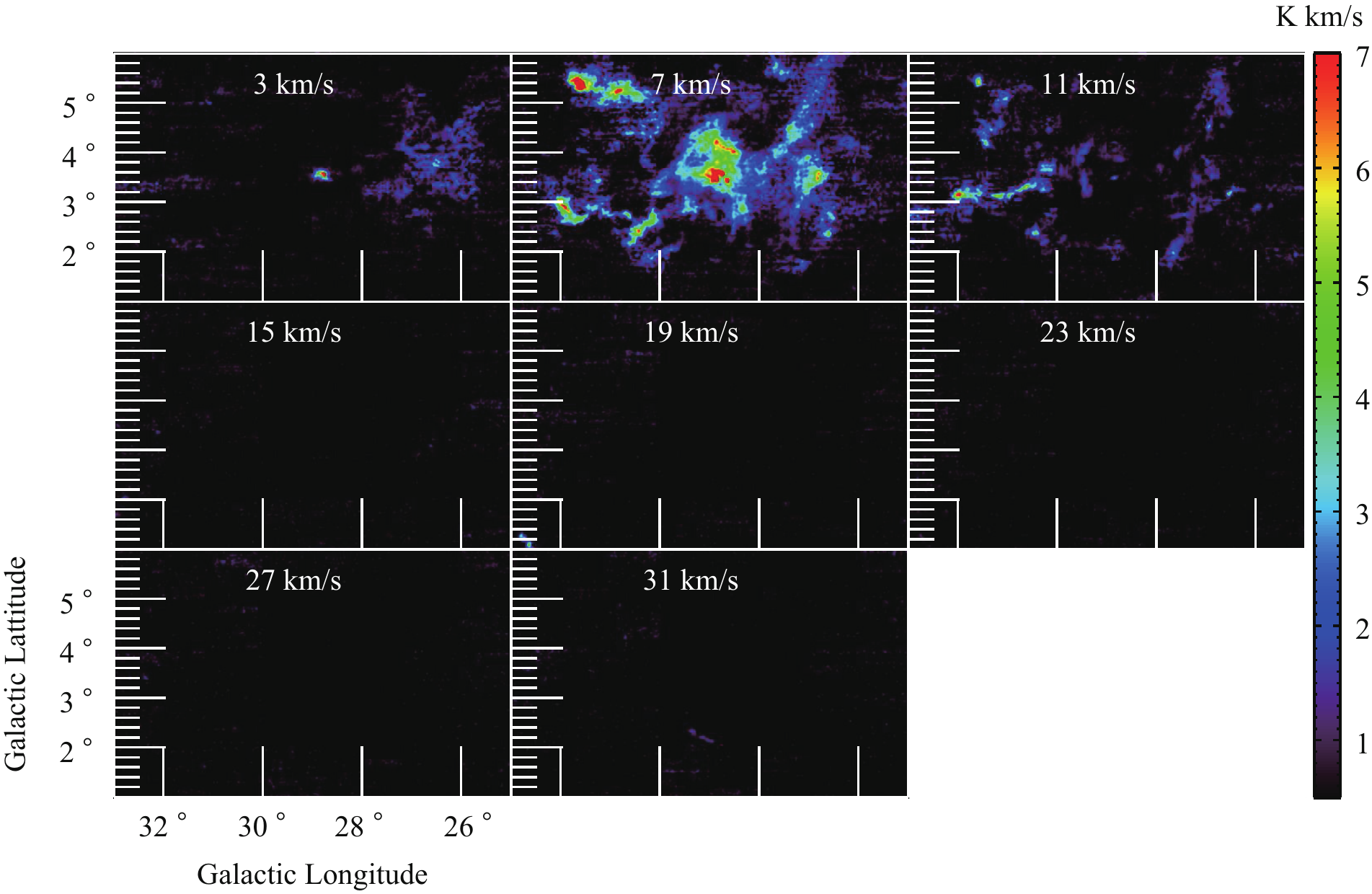}
\caption{
$^{13}$CO ($J=2-1$) velocity channel maps
toward the Aquila Rift and Serpens molecular cloud complexes.
(Animations of the channel maps are available in the online journal.)}  
\label{fig:13CO channel map}
\end{figure}

\begin{figure}
\plotone{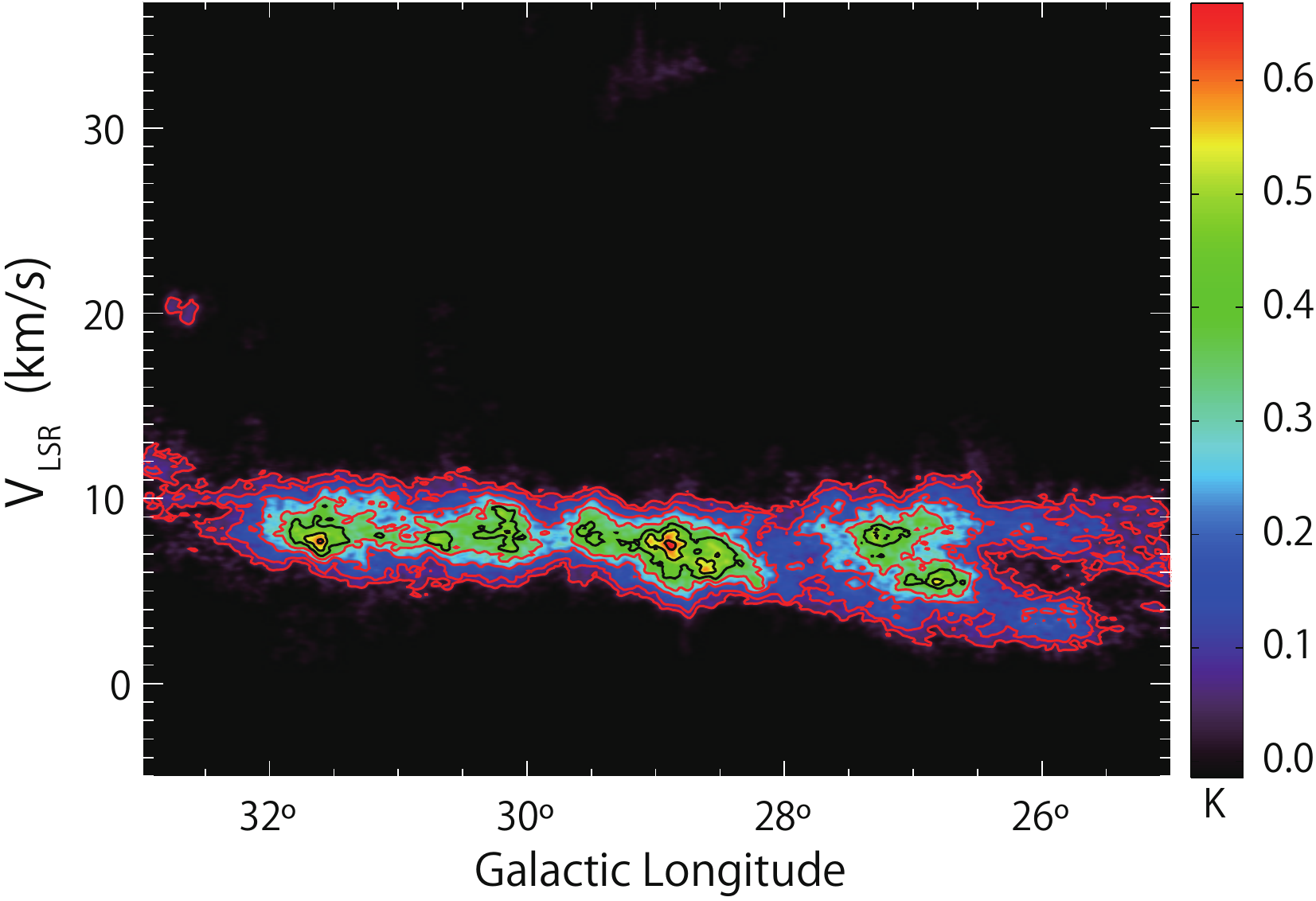}
\caption{
$^{13}$CO ($J=2-1$) longitude-velocity diagram integrated over the entire
 Galactic latitude range observed ($1^\circ < b < 6^\circ$).
 The contours are drawn at 0.05, 0.1, 0.2, 0.4, 0.5, and 0.6 K.
}
\label{fig:PV map2}
\end{figure}

For comparison, we show $^{13}$CO ($J=2-1$)  velocity integrated intensity map toward the Aquila Rift and Serpens molecular cloud complexes in Figure \ref{fig:13CO integ map}.   
The $^{13}$CO ($J=2-1$) channel maps and the 
longitude-velocity diagram integrated over the Galactic latitude 
range of $1^\circ \le b \le 6^\circ$ are indicated in Figures \ref{fig:13CO channel map} and  \ref{fig:PV map2}, respectively.
Comparison between the $^{13}$CO map and the extinction map indicates that 
the $^{13}$CO emission better traces the extinction image than the $^{12}$CO emission 
(Figure \ref{fig:13CO integ map}b).
Two components with different velocities seen in the $^{12}$CO channel maps are also recognized in the $^{13}$CO channel maps.
Some arcs identified with $^{12}$CO can be seen in the $^{13}$CO map, indicating that densities of those arcs are relatively high.
The longitude-velocity diagram shows that only 8km s$^{-1}$ component
is prominent in the $^{13}$CO map.

\subsection{Comparison of Molecular Gas and Visual Extinction}
\label{sec:av}

Here we compare the CO data with dust detected in the $E(J-H)$ map.
While the CO lines are emitted only in dense regions
(e.g., $\gtrsim 10^{2-3}$ cm$^{-3}$),
the color excess map traces the total dust column density
along the line of sight including diffuse regions in the background
(or foreground) unrelated to the CO emitting volumes.
We therefore should remove the background of
the $E(J-H)$ map to compare the two dataset directly.

For compact clouds at high latitudes, such removal of the background
can be easily done by subtracting a constant or by fitting the background
with an exponential function of the Galactic latitudes $b$ 
\citep[e.g.,][]{dobashi05}.
Those methods, however, do not work well in the case of the clouds in Aquila Rift,
because the region suffers from great complexity over a large extent.
We therefore decided to use the following procedure to remove the background:
We first defined the temporary background as the regions without showing apparent small 
scale-structures in Figure \ref{fig:2mass}(a), and masked the apparent clouds in the figure by eye-inspection. 
We then fitted the unmasked pixels by two dimensional (2D) polynomial function of $N=4$ degree,
and subtracted it from the original $E(J-H)$ map to convert the residual map to $A_V$ as
\begin{equation}
A_V = 9.35 E(J-H)  \ ,
\label{eq:conversion_to_Av} 
\end{equation}
where the coefficient is calculated using the reddening law
found by \citet{reike85}. 
We smoothed the resulting $A_V$ map with $10\times 10$ pixels ($=10'\times 10'$), and redefined the background
as the regions with $A_V <0.5$ mag in the smoothed map. The background regions defined
in this manner is shown in Figure\ref{fig:2mass} (b). 
To better assess the background,  we further fitted
the background regions in the figure with a 2D polynomial function with higher orders ($N$),
and decided to adopt $N=8$ degree to produce the final $A_V$ map which we use in the following 
analyses. We show the map in Figure \ref{fig:2mass}(c).
The maximum value found in the map is $A_V\simeq 50$ mag,
and the $1\sigma$ noise level of the map is $\delta A_V \simeq 2.5$ mag.

In order to infer the possible systematic error due to the background removal,
we reproduced background maps with different orders of polynomials $N=4-9$
to derive $A_V$ maps in the same way as described above.
We found that the $A_V$  values in the resulting maps vary by $\sim3$ mag at most
depending on $N$, which we regard as the uncertainty in our $A_V$ map
arising from the background determination. 

\begin{figure}
\plotone{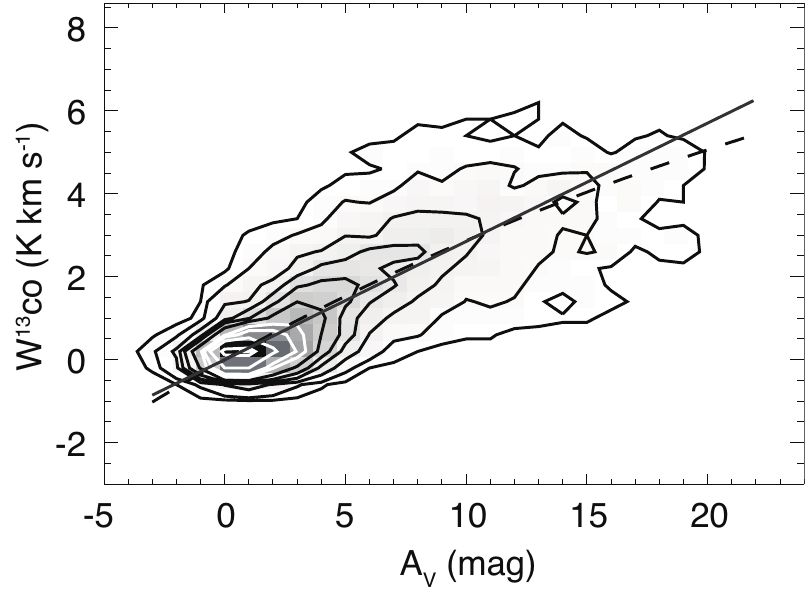}
\caption{The $^{13}$CO integrated intensity versus $A_V$.
The  solid and broken lines represent
Equations  (\ref{eq:linear_13CO}) and (\ref{eq:exp_13CO})
best fitting the relations, respectively.
Contours are drawn to exclude 1, 5, 15, 25, 35, 45, 55, 65, 75, 85, and 95 $\%$
of the total data points (i.e., 99$\%$ of the data points 
are included within the lowest contour).
}  
\label{fig:13co_vs_Av}
\end{figure}

\begin{figure}
\plotone{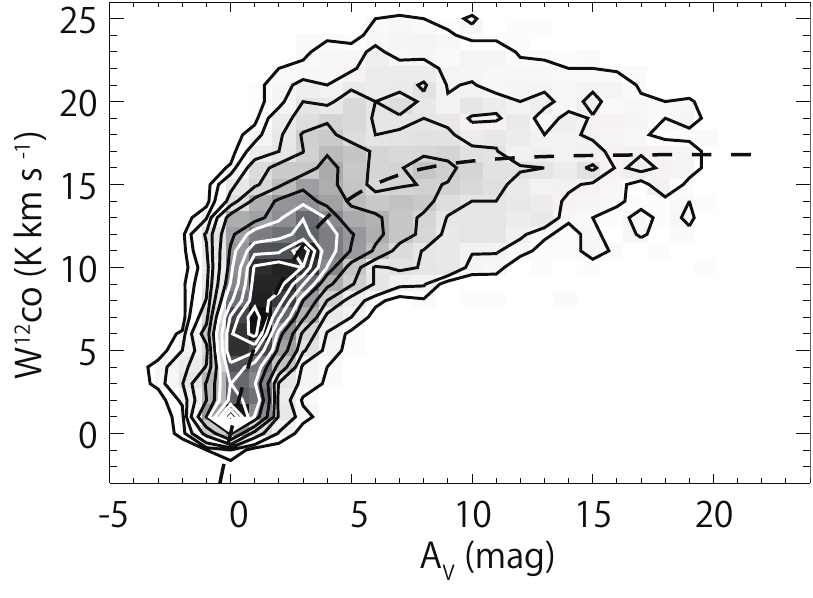}
\caption{The $^{12}$CO integrated intensity versus $A_V$.
The relation is shown in the plot density.
The broken line represents
Equation (\ref{eq:exp_12CO}) best fitting the relation.
Contours are the same as in Figure \ref{fig:13co_vs_Av}.
}  
\label{fig:12co_vs_Av}
\end{figure}

To compare the CO and  $A_V$ data directly, we smoothed both of the data
to the common 6$'$ (FWHM) angular resolution. We excluded some positions from the
analyses where the $A_V$ map suffers from a very low angular resolution ($> 6'$).
Figures \ref{fig:13co_vs_Av} and \ref{fig:12co_vs_Av} show the relations of
the $^{13}$CO and $^{12}$CO integrated intensities vs. $A_V$, respectively.
In the case of the $^{13}$CO emission, the relation is simple and can be fitted by a linear function,
while that of the $^{12}$CO emission is complex mainly due to the heavy saturation at higher $A_V$.
This saturation is presumably due to the fact that $^{12}$CO emission becomes
optically thick at higher $A_V$.

In order to estimate the total masses of clouds that we identified
with the molecular data (see Section \ref{sec:clouds}), we formulate the
dependence of $W_{\rm ^{13}CO}$ on $A_V$.
For this purpose, we fitted the  $W_{\rm ^{13}CO}$ vs. $A_V$  relation 
by a linear function, 
\begin{equation}
W_{\rm ^{13}CO} = {a_0}{A_V} \ ,
\label{eq:linear_13CO} 
\end{equation}
where the coefficient best fitting the relation is
\begin{equation}
a_0=0.285 \pm 0.001 \ {\rm K \ kms^{-1} mag^{-1}}  \ .
\end{equation}

The relations of $W_{\rm ^{12}CO}$ vs. $A_V$ and $W_{\rm ^{13}CO}$ vs. $A_V$ 
appear to be better fitted with exponential functions.
Here, we attempt to fit these relations with exponential functions. 
Similar attempt has been done by \citet{pineda10}.

For the $W_{\rm ^{13}CO}$ vs. $A_V$  relation, we consider  
the following exponential function,
\begin{equation}
W_{\rm ^{13}CO} = {a_1}\left[ {1 - \exp ( - {a_2}{A_V})} \right]   \ ,
\label{eq:exp_13CO} 
\end{equation}
where $a_1$ and $a_2$ are constants.

The coefficients best fitting the relation are obtained as
\begin{equation}
a_1=12.16 \pm 0.55 \ {\rm K \ kms^{-1}}, 
\end{equation}
and
\begin{equation}
a_2=0.0269 \pm 0.014 \ {\rm mag^{-1}} .
\end{equation}

For the $W_{\rm ^{12}CO}$ vs. $A_V$  relation, 
we consider the following exponential function
\begin{equation}
W_{\rm ^{12}CO} = {b_1}\left[ {1 - \exp ( - {b_2}{A_V})} \right]  , 
\label{eq:exp_12CO} 
\end{equation}
where $b_1$ and $b_2$ are constants.

The coefficients best fitting the relation are derived as
\begin{eqnarray}
b_1 &=& 16.8 \pm 4.4 {\rm K \ km \ s^{-1}} , \\
&& {\rm and} \\
b_2&=& 0.37 \pm 0.18 \ {\rm mag^{-1}} .\\
\end{eqnarray}

The relations expressed by Equations (\ref{eq:exp_13CO}) and (\ref{eq:exp_12CO})
are shown in Figures \ref{fig:13co_vs_Av} and \ref{fig:12co_vs_Av}, respectively.

The $X_{\rm CO}$ factor gives a scaling between CO luminosity and molecular cloud mass and we use $X_{\rm CO}$ to derive the masses of the clouds 
identified in the next section.
As shown above, $W_{\rm ^{13}CO}$ is well correlated with $A_V$.
Therefore, we use $^{13}$CO emission to identify the clouds and evaluate their masses. 
The $X_{\rm ^{13}CO}$ factor is calculated as
\begin{equation}
X_{\rm ^{13}CO} \equiv {N_{\rm H_2} \over W_{^{13}CO}}= 3.3 \times 10^{21} {\rm cm^{-2} K^{-1} km^{-1} s} ,
\label{eq:xfactor}
\end{equation}
where we used Equation (\ref{eq:linear_13CO}) and assumed the relation of
$N({\rm H_2})/A_V = 9.4 \times 10^{20} $ cm$^{-2}$ mag$^{-1}$ and $R_V = 3.1 $ \citep{bohlin78}.

Adopting the above $X_{\rm ^{13}CO}$ factor, the total molecular gas mass in the observed
region is estimated to be about $ 1 \times 10^5 M_\odot$ under the assumption of $d=260$ pc.

The $X_{\rm CO}$-factors are usually estimated by using the $^{12}$CO ($J=1-0$) line.
For our Galaxy, the averaged $X_{\rm ^{12}CO}$ is evaluated to be 
$ 2 \times 10^{20}$ cm$^{-2}$ K$^{-1}$ km$^{-1}$ s with an uncertainty of $\pm 30 \%$
\citep{bolatto13}.  This values vary from region to region.
For example, \citet{pineda10} derived the $X$-factor of $(1.6 \sim 12) \times 10^{20}$ cm$^{-2}$ K$^{-1}$  km$^{-1}$ s 
in the Taurus molecular cloud. 
For L1551, \citet{lin16} measured the $X_{\rm CO} = 1.08 \times 10^{20}$ cm$^{-2}$ K$^{-1}$ km$^{-1}$ s.

Here, we attempt to compare our $X$-factor using higher transition with these previous values using the $J=1-0$
line.
We choose the Serpens Main Cloud because this region basically has single component. In constrast,
the Serpens South+W40 region contains multiple-components.  
The part including Serpens Main Cloud has $W_{\rm ^{12}CO} = 20 - 25$  K km s$^{-1}$ and $W_{\rm ^{13}CO} = 4 - 6$ K km s$^{-1}$.  
In the part including Serpens Main Cloud,  the mean intensity of $^{12}$CO $(J=2-1)$ and line width are evaluated to be 5.6 K and 3.5 km s$^{-1}$, respectively. 
The mean intensity of $^{13}$CO ($J=2-1$) and line width are evaluated to be 2.5 K and 2.5 km s$^{-1}$, respectively. 
Adopting the H$_2$ density of $10^3$ cm$^{-3}$, $T=10$ K, the H$_2$ column density of $1\times 10^{21}$ cm$^{-2}$  ($A_V \approx 15-20$), 
the CO fractional abundance of $10^{-4}$, and $^{12}$CO/$^{13}$CO ratio
of 80,  the intensity of $^{12}$CO ($J=1-0$) is derived as $\sim$ 7 K using \texttt{RADEX} \citep{vandertak07}, and thus $W_{\rm 12CO (1-0)} = 25$ K km s$^{-1}$.  
Thus, we can convert our  $X_{\rm 13CO (2-1)}$ to $X_{\rm 12CO (1-0)}= 5.6 \times 10^{20}$ cm$^{-2}$ K$^{-1}$  km$^{-1}$ s,
which is about 2.5 times larger than the standard value of \citet{bolatto13}.  
Therefore, our $X_{\rm 13CO (2-1)}$ value is consistent with the previous estimation, taking into account a large variation of $X$-factor values.

\section{Identification of Molecular Clouds}
\label{sec:clouds}

As shown in Section \ref{sec:results}, the correlation between 
$^{13}$CO integrated intensity and extinction $A_V$ is reasonably good.
Therefore, we identify molecular clouds from the $^{13}$CO data cube
and derive their physical quantities such as cloud masses, radii, and line widths.

There are several methods proposed to identify clouds.
In the following, we apply a new method called \texttt{SCIMES} \citep{colombo15} 
to our CO data cube. \texttt{SCIMES} uses 
the output of \texttt{Dendrogram} \citep{rosolowsky08}, 
and identify the structures that are clustering, based on a graph theory.

\subsection{\texttt{Dendrogram} and \texttt{SCIMES} Analysis}

\texttt{Dendrogram} characterizes the hierarchical structure of the isosurfaces 
for molecular line data cubes and have been used to identify the structures
like cores and clumps \citep{rosolowsky08}.
However, it is sometimes difficult to characterize the structures identified
only with \texttt{Dendrogram}.  Recently, \citet{colombo15} constructed a new method 
called \texttt{SCIMES} (Spectral Clustering for Interstellar Molecular Emission 
Segmentation) to identify clustered structures with similar emission 
properties as "clouds" based on the hierarchical structures 
identified by \texttt{Dendrogram}.  
Applying \texttt{SCIMES} to the $^{12}$CO ($J=1-0$) data cube of Orion A,
\citet{colombo15} demonstrated that \texttt{SCIMES} can identify 
the cloud structure well.  In the following, we  apply \texttt{SCIMES} to our
$^{13}$CO data cube because $^{13}$CO trace the cloud structures better than $^{12}$CO
(see the previous section) and call a clustered structure identified by \texttt{SCIMES} as a "cloud".
Before applying \texttt{Dendrogram} to our data, we masked pixels whose intensities are about twice the
noise level of each observation box.

\begin{figure}
\plotone{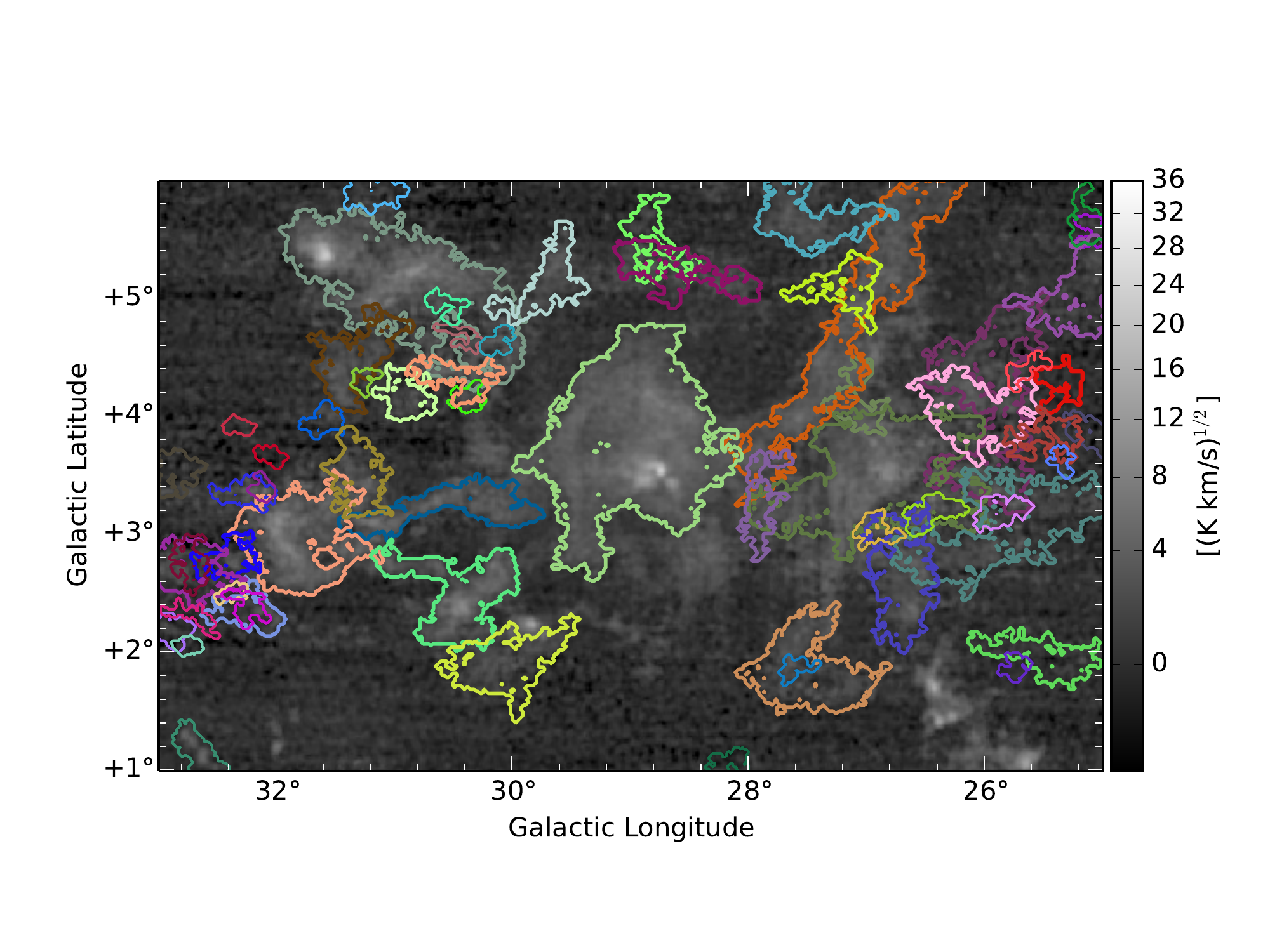}
\caption{Clouds identified by \texttt{SCIME} overlaid on the $^{13}$CO integrated intensity map. 
The clouds identified are enclosed by colored curves. 
}  
\label{fig:scime}
\end{figure}

\texttt{Dendrogram} requires three input parameters:
\texttt{min\_value}, \texttt{min\_delta}, and \texttt{min\_npix}.
The first parameter, \texttt{min\_value}, specifies the minimum value above which 
\texttt{Dendrogram} identifies the structures.  The second parameter, \texttt{min\_delta},  is the minimum height 
required for a structure identified. The third parameter, \texttt{min\_npix}, is 
the minimum number of pixels that a structure should contain in order to remain 
an independent structure.
We apply \texttt{Dendrogram} with the following three input parameters
of \texttt{min\_value} = 2 $\sigma$, \texttt{min\_delta} = 3 $\sigma$, and \texttt{min\_npix} = 50,
where $\sigma$ is the rms noise level of the data.
The value of  \texttt{min\_npix} = 50 is adopted to estimate the physical quantities 
with reasonable accuracy.
For the \texttt{Dendrogram} analysis, we adopt $\sigma = 0.38$ K.  
The \texttt{Dendrogram} classifies two types of structures: {\it leaves} and {\it branches}.
{\it Branches} are the structure which split into multiple sub-structures, and {\it leaves} are the structures 
which do not have any sub-structures.  
See \citet{rosolowsky08} for the detail of the \texttt{Dendrogram} analysis.  
Then, we applied \texttt{SCIMES} with the results of \texttt{Dendrogram}.
We use the element-wise multiplication of the luminosity and the volume matrix for analyzing the clustering of the dendrogram.  See \citet{colombo15} for more details.
In total, we identified 61 clouds. We classified all the structures as clouds.  
In Figure \ref{fig:scime}, we also plot the clouds identified on the $^{13}$CO velocity-integrated intensity map.
The structure enclosed by each colored curve corresponds to the cloud identified.

In Table 2, we list positions and some physical quantities of the clouds identified such as 
the cloud velocity $V_{\rm LSR}$, the lengths of major and minor axes, and position angles.
The positions of a cloud are determined as the intensity-weighted positions of the structure 
identified in the corresponding directions.
Major and minor axes of the projected structure onto the Galactic longitude-latitude plane, $R_{\rm maj}$ and
$R_{\rm min}$,  are computed from the intensity-weighted second moment in direction of greatest elongation 
and  perpendicular to the major axis in the Galactic longitude-latitude plane, respectively.
The position angle is the angle of the major axis in degrees counter-clockwise 
from the longitude axis.
The region including the W40 H\,{\sc ii} region and Serpens South embedded cluster is
assigned as a single cloud. The Serpens Main cloud is also identified
as a single cloud.

\subsection{Physical properties of the identified clouds}

\subsubsection{Derivation of Physical Quantities}

In Table 3, we list cloud radii, masses, and line widths  of the clouds identified.
The cloud radii are computed as $1.91 \times \sqrt{R_{\rm maj} R_{\rm min}}$, following the 
definition of \citet{rosolowsky08}.
The cloud velocities are calculated as the intensity-weighted local standard of rest (LSR) velocity 
of the structure identified. The line widths are the intensity-weighted FWHM values.
The cloud mass is computed using the relation 
\begin{equation}
 M = \sum I_{ijk} \Delta V_j \Delta x_i \Delta y_j /X_{\rm ^{13}CO} .
\end{equation}
where $I_{ijk} $ is the intensity in the brightness temperature scale at the $i$, $j$, $k-$th grid in the position-position-velocity space. 
The spacings of $\Delta x_i$ and $\Delta y_j$ are the sizes of each grid in the Galactic longitude and latitude directions, respectively,
and $\Delta x_i = \Delta y_j = 60" \times 260 \ {\rm pc} \times 1.5\times 10^{13} $ 
$=2.3 \times 10^{17}$ cm.
The velocity width of $\Delta V_k$ is the velocity difference of adjacent channels and $\Delta V_k = 0.083$ km s$^{-1}$.

\begin{figure}
\plotone{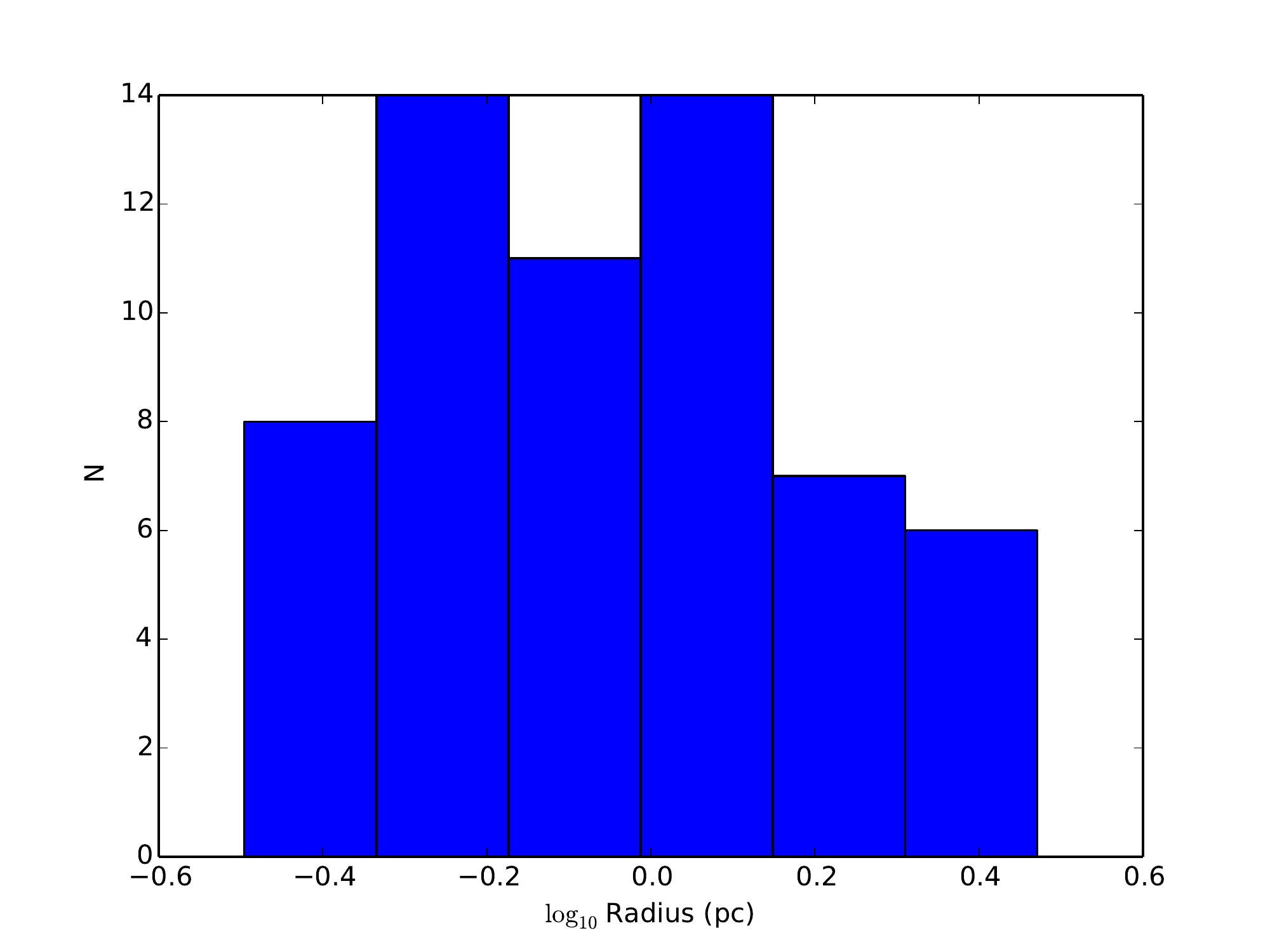}
\caption{Histogram of Cloud Radius.
}  
\label{fig:radius}
\end{figure}

\begin{figure}
\plotone{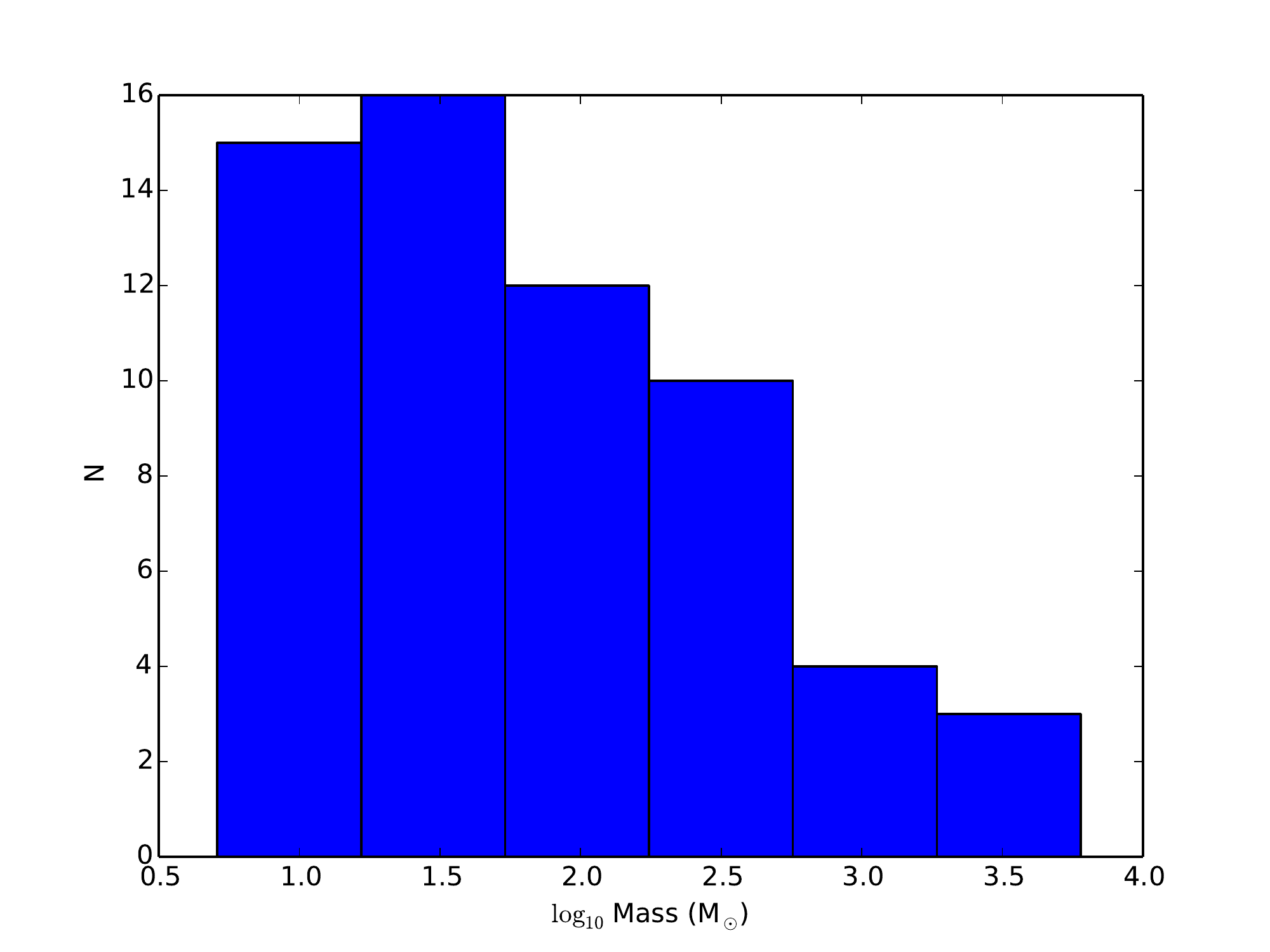}
\caption{Histogram of Cloud Mass.
}  
\label{fig:mass}
\end{figure}

\begin{figure}
\plotone{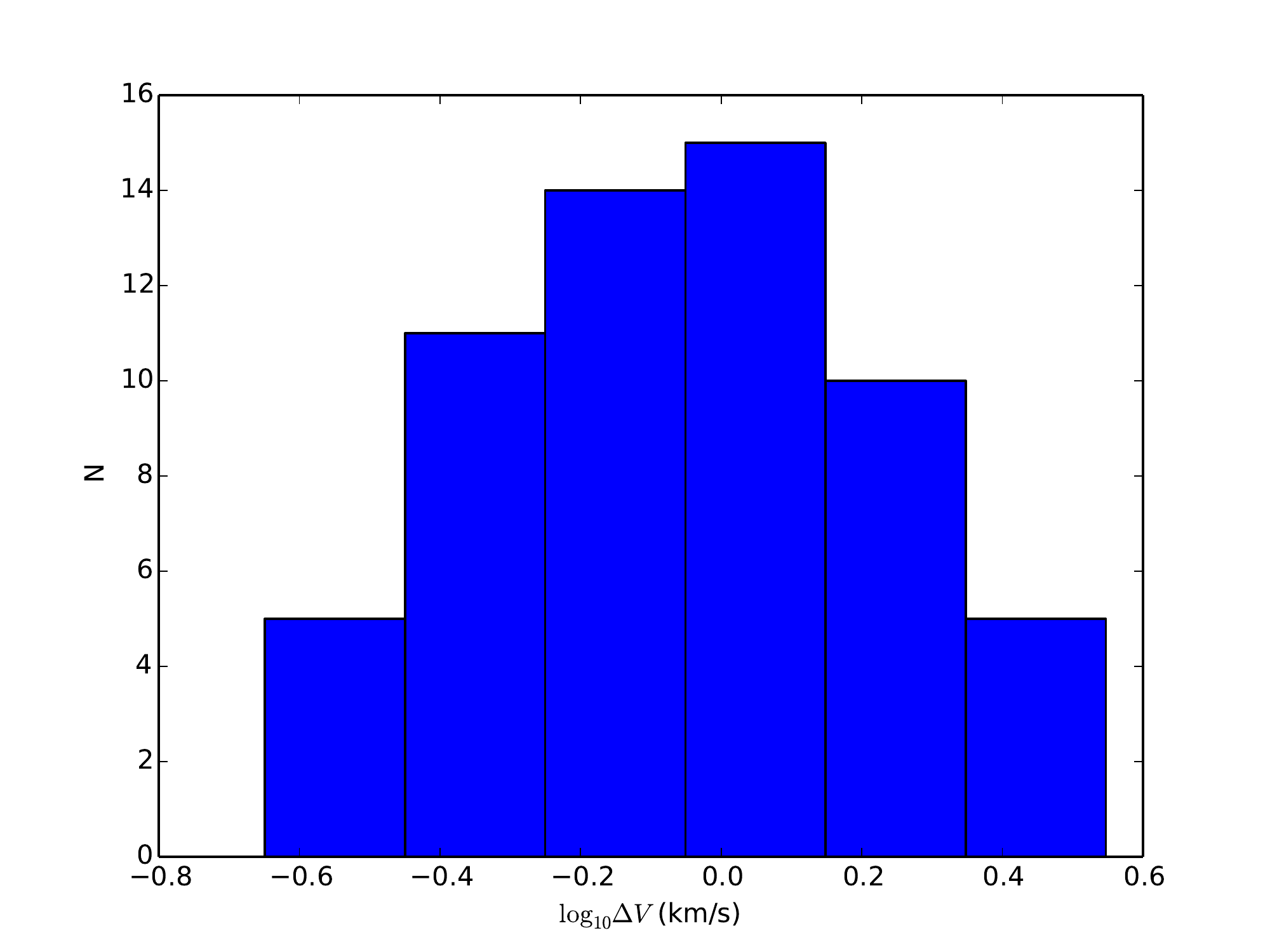}
\caption{Histogram of Line Widths.
}  
\label{fig:linewidth}
\end{figure}

Figures \ref{fig:radius}, \ref{fig:mass}, and \ref{fig:linewidth} show the histograms of the radius, mass, and
line width of the identified clouds, respectively.
The radii of the identified clouds range from 0.3 to 3 pc and its mean value is around 
$\sim$ 1 pc.  
The cloud masses range from 5 $M_\odot$ to $6\times 10^3 M_\odot$. 
 The distribution of cloud masses has a peak at around 30 $M_\odot$, 
which indicate that there are many less-massive clouds. 
Here, we assumed $d=260$ pc.
The most massive cloud has about $6\times 10^{3} M_\odot$ and contains 
the Serpens South and W40 region, which are located near the center of the observation box.
This cloud has at least two different components with different velocities
(The velocities of the two peaks are 5.9 and 7.6 km s$^{-1}$), but it
is classified as a single cloud (No. 16 cloud with $V_{\rm LSR}=7.1$ km s$^{-1}$).
The second most massive cloud is a molecular cloud which contains the Serpens Main region 
or Serpens Cloud Core (Serpens cluster A and B in Figure \ref{fig:2mass})
A Herbig Ae Be protostellar system EC 95 is
 located in Serpens Cloud Core whose line-of-sight velocity is 
 measured as
 $V_{\rm LSR}= 8-9$ km s$^{-1}$ \citep{mcmullin00}. 
The distance of EC95 is measured from the VLBA observations as 415 pc.
 In Table 2,  the Serpens Cloud Core is associated with No. 2 cloud with $V_{\rm LSR}=8$ km s$^{-1}$.
If EC95 is really associated with No. 2 cloud, the distance of No. 2 cloud may be almost 415 pc.
The third most massive cloud is associated with MWC 297.  This has the largest line width.
This is identified as No. 9 cloud with   $V_{\rm LSR}=5.4$ km s$^{-1}$.
The total mass is estimated to be about $2.2\times 10^4 M_\odot$.
The line widths ranges from $\sim$ 0.3 to 4 km s$^{-1}$ 
 and its distribution has a single peak at around 1 km s$^{-1}$.
The cloud identified have basically supersonic internal motions.

\subsubsection{Line-width-radius Relation}

\begin{figure}
\plotone{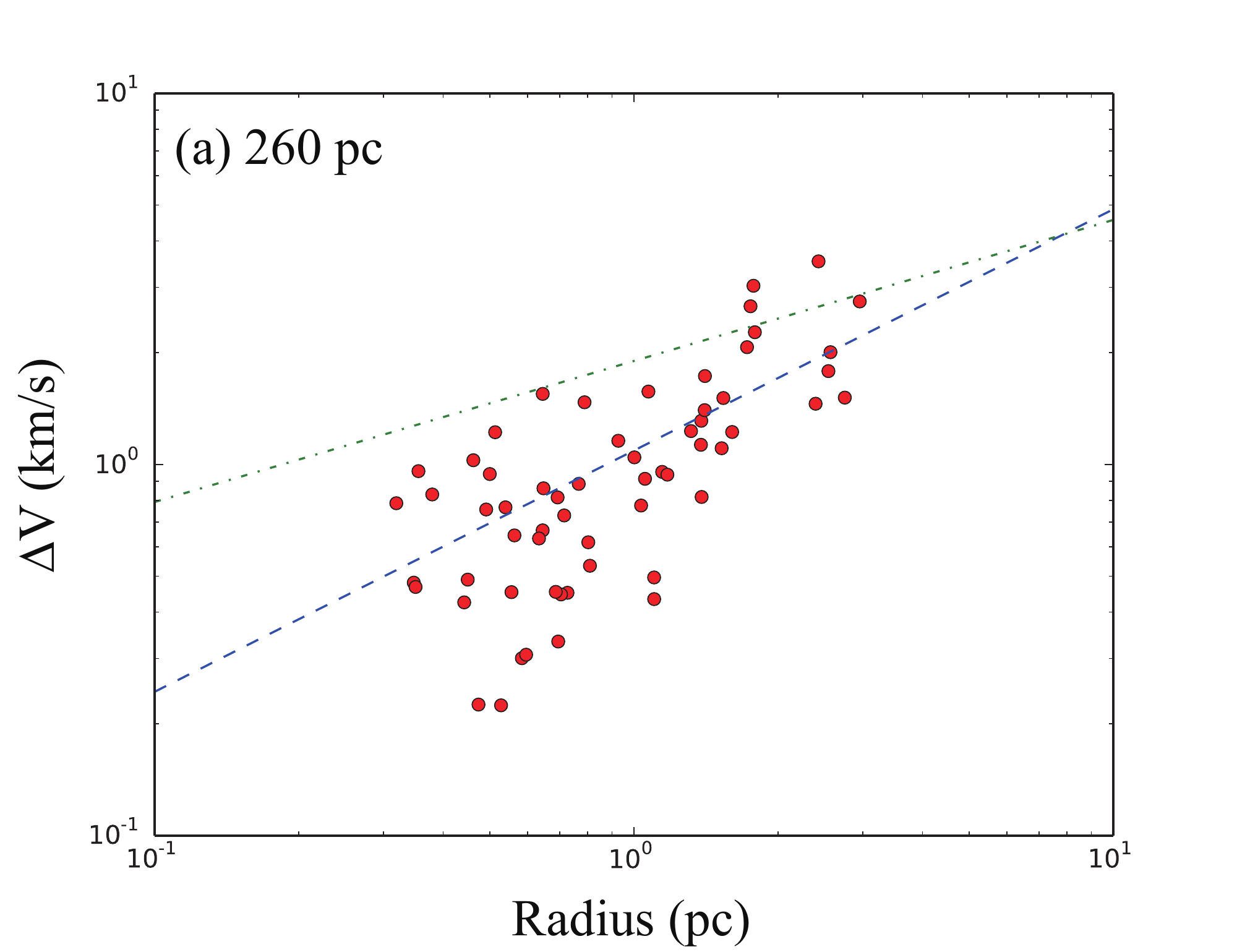}
\plotone{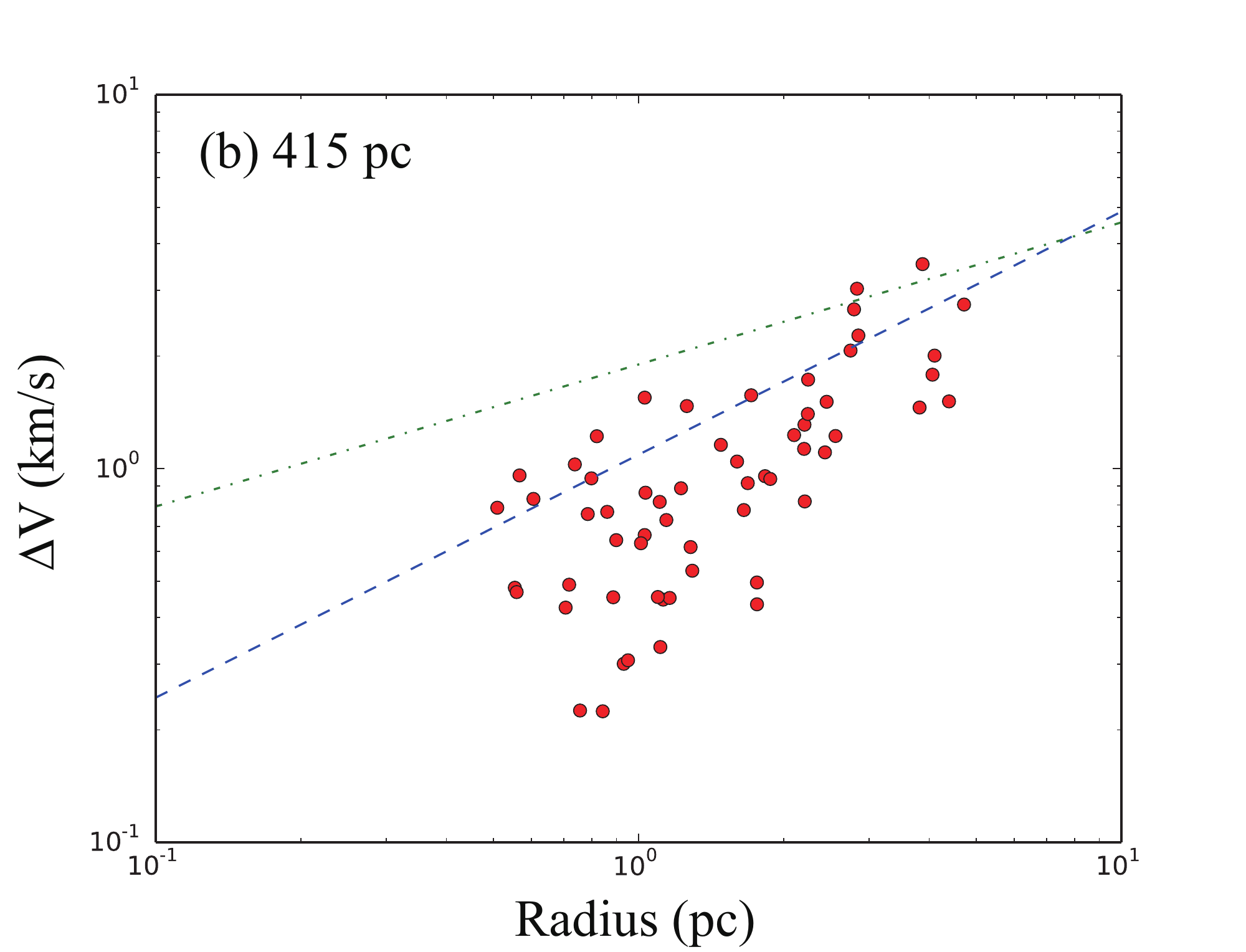}
\caption{Line-width-Radius Relations assuming a distance of (a) 216 pc and (b) 415 pc. 
The dashed and dashed-dotted lines are the line-width-radius relations derived by 
\citet{heyer04} and \citet{larson81}, respectively.} 
\label{fig:line-width-radius}
\end{figure}

Line widths of the clouds provide us with information of internal cloud turbulence.
In Figure \ref{fig:line-width-radius}a, we present the line width-radius relation of the identified cloud.  
Here, we assumed a diatance of 260 pc.
For comparison, we show the line width-radius relations obtained by 
\citet{heyer04}, \citet{shetty12}, \citet{colombo15}, and \citet{larson81}. 
For the relations of \citet{heyer04} and \citet{larson81},  we modified the 
original relations presented in the corresponding papers so as to match 
the definitions of the radii and line widths.  See the appendix of
\citet{maruta10} for the details. 
It is worth noting that \citet{shetty12} and \citet{colombo15} used 
\texttt{Dendrograms} and \texttt{SCIMES} to identify the clouds, respectively,
and therefore can be directly compared to our result.
The two line-width-size relations of \citet{shetty12} are derived based on the $^{13}$CO
($J=1-0$) data of Perseus molecular cloud and N$_2$H$^+$ ($J=1-0$) data of the 
Central Molecular Zone (CMZ).
The line-width-size relation of \citet{colombo15} is derived based on the $^{12}$CO
($J=1-0$) data of Orion A molecular cloud.
The cloud line-width of Perseus by \citet{shetty12} is about 1.5 times larger 
than that of Orion A by \citet{colombo15} at a given radius. 
 Our line-width-radius relation is consistent with the \citet{heyer04} relation,  
 Perseus relation by \citet{shetty12}, and Orion A relation by \citet{colombo15}.
If we adopt a distance of 415 pc, the line-widths of the clouds tend to be smaller than 
those of \citet{heyer04}'s Perseus and Orion A relations.  
For comparison, we show the line-width-radius relation of the identified clouds
in Figure \ref{fig:line-width-radius}b, where we assumed $d= 415$ pc.
For both the distances, the clouds in CMZ have significantly larger line-widths, 
which support the results of \citet{shetty12}.

\subsubsection{Virial-Parameter-Mass Relation}

\begin{figure}
\plotone{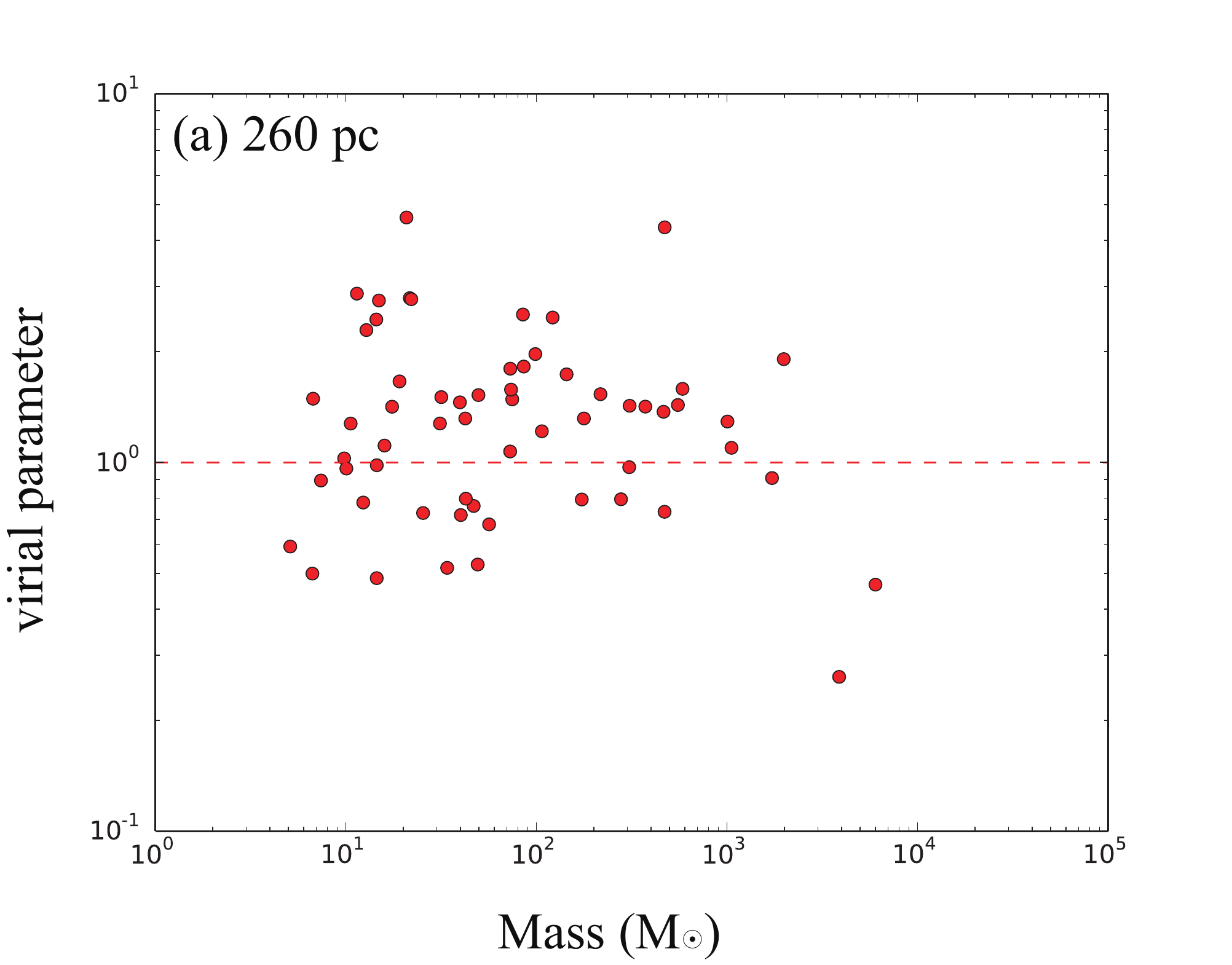}
\plotone{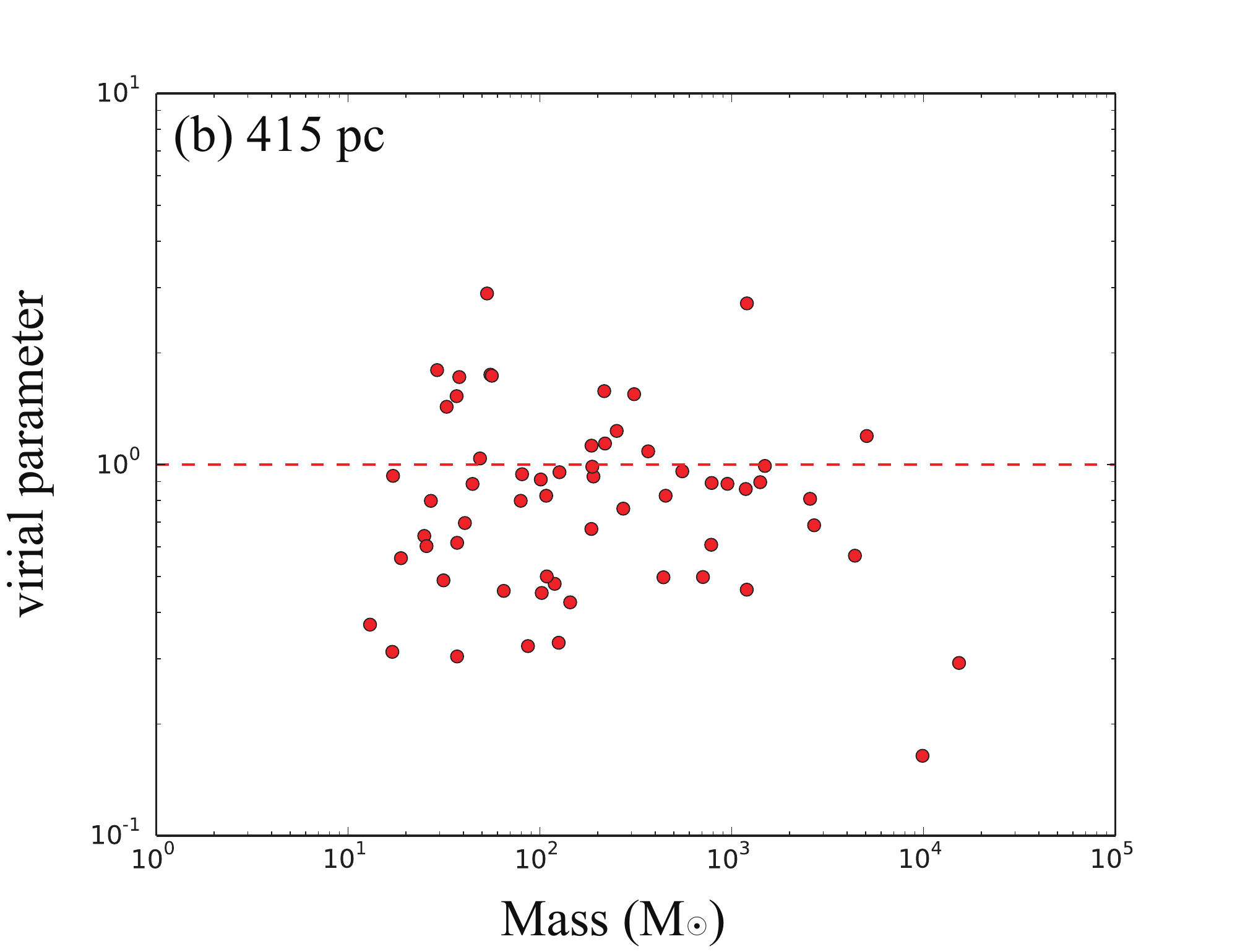}
\caption{Line-width-Radius Relations  assuming a distance of (a) 216 pc and (b) 415 pc. 
The dashed lines indicate virial equilibrium.}  
\label{fig:virial-mass}
\end{figure}

The gravitational boundedness is often evaluated by the virial parameter.
In Figure \ref{fig:virial-mass}a, we plot the virial parameter against the cloud mass. 
Here, we define the virial parameter as
\begin{equation}
\alpha _{\rm vir} = {5 a^{-1} R_{\rm cl} \Delta V^2 \over 8 \ln 2 G M_{\rm cl}} \ ,
\end{equation}
where $a$ is constant and we set $a=5/3$ corresponding to a centrally-condense
sphere with $\rho \propto r^{-2}$.  For uniform sphere, $a=1$.
We did not apply the correction of the thermal line width of the H$_2$ gas to the line width 
$\Delta V$ because its contribution to the total line width should be small as long as 
the gas temperature stays at around 10 K.
Majority of the clouds identified has virial parameters close  to unity.   
Even for smaller clouds, the virial parameters are close to or less than the unity.
This indicates that most of the clouds identified are gravitationally-bounded. 
It is worth noting that similarly to the line-width-radius relation, the virial-parameter-mass 
relation is affected 
by the adopted distance. If we adopt a distance of 415 pc, the clouds
tend to become more gravitationally-bounded.
In either way, the majority of the clouds are gravitationally-bounded both for $d\sim 260$ pc and 415 pc. 
This  may be the reason why the northern part of Aquila Rift contains several active star-forming regions.
We speculate that interaction of superbubbles, large-scale turbulent flows and/or cloud-cloud collisions 
may have trigger star formation in this region.
Accurate distance measurements are crucial to clarify the dynamical states of the clouds in this region.

\subsubsection{Mass-Radius Relation}

\begin{figure}
\plotone{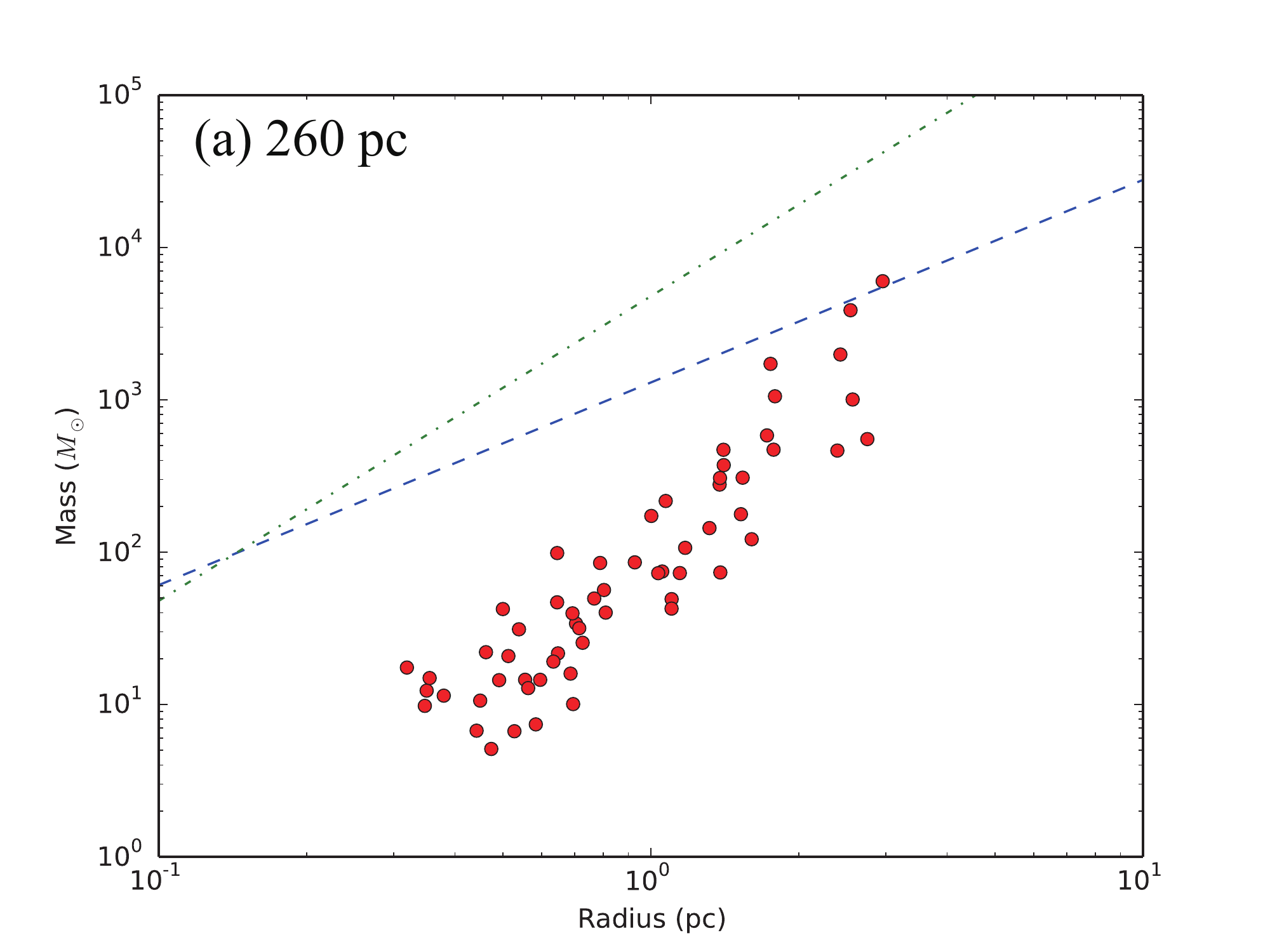}
\plotone{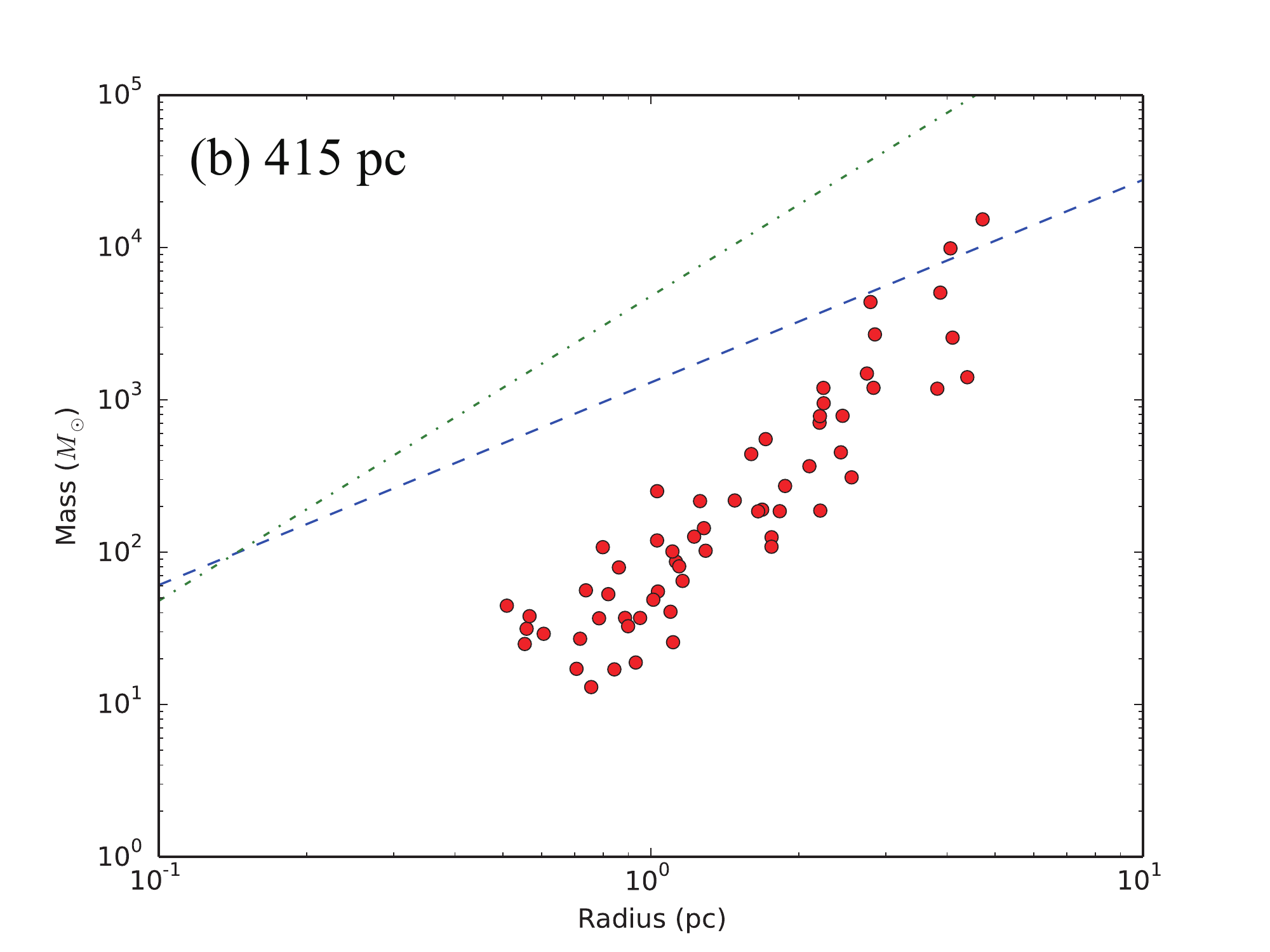}
\caption{Mass-Radius Relations  assuming a distance of (a) 216 pc and (b) 415 pc. 
The dashed and dashed-dotted lines indicate \citet{kauffmann10}'s relation and \citet{krumholz08}'s criteria
for massive star formation, respectively.}  
\label{fig:massradius}
\end{figure}

Finally, we present the mass-radius relation of the identified clouds in Figure \ref{fig:massradius}, where
we adopt a distance of 260 pc and 415 pc for the panels (a) and (b), respectively.
According to \citet{kauffmann10}, the clouds forming massive stars ($M \gtrsim 10M_\odot$) 
have masses larger than $M  \gtrsim 1300 M_\odot (r/{\rm pc})^{1.33}$, 
where we rescaled  \citet{kauffmann10}'s relation by a factor of 1.5 to take into account the difference
in the adopted dust parameters, following \citet{tanaka13}. In addition, \citet{krumholz08} derived 
the threshold column density of 1 g cm$^{-2}$ for massive star formation.
For comparison, we indicate \citet{kauffmann10}'s relation and \citet{krumholz08}'s criteria 
for massive star formation in Figure \ref{fig:massradius} with dashed and dashed-dotted lines, respectively.
For both panels, most of the identified clouds are distributed below the two lines.
Most massive clouds are located near \citet{kauffmann10}'s line.
Therefore, we suggest that in the observed area, it may be difficult to form massive stars in near future
without any external events such as cloud-cloud collision which can increase the local column densities. 
In fact, there are only a couple of regions associated with H\,{\sc ii} regions in the observed area,
W40 and MWC 297.  The exciting stars of the H\,{\sc ii} regions are late O-type and B-type stars 
with masses of $\approx 10 M_\odot$.
This fact supports the idea that the Aquila Rift and Serpens cloud complexes may not be 
a massive star-forming region.

\section{Summary}
\label{sec:summary}

We summarize the main results of the present paper as follows.

\begin{itemize}
\item[1.] We carried out large-scale $^{12}$CO ($J=2-1$), $^{13}$CO ($J=2-1$), and C$^{18}$O ($J=2-1$) 
mapping observations toward 
the Aquila Rift and Serpens region of $ 25^\circ < l < 33^\circ$ and 
$1^\circ < b < 6^\circ$ at a effective angular resolution of 3$'$.4 ($\approx 0.25$ pc) 
and at a velocity resolution of $\sim$ 0.08 km s$^{-1}$ 
with the velocity coverage of $-5$ km s$^{-1} <$ $V_{\rm LSR} < $  35 km s$^{-1}$.

\item[2.] From the CO channel maps, we found a number of arcs, which extend
over 1 $-$ 7 pc. These structures may have been formed by the superbubbles that
converge toward our observed area \citep[see also][]{frisch98}. 

\item[3.] The velocity integrated intensities of $^{13}$CO are well correlated to 
the 2MASS visual extinction.  We derived the $X_{\rm ^{13}CO}$ factor,
\begin{equation}
X_{\rm ^{13}CO} =  3.3 \times 10^{21} {\rm cm^{-2} K^{-1} km^{-1} s} \ ,
\end{equation}
 by comparing the $^{13}$CO velocity integrated intensity map with the 2MASS extinction map.
 
 \item[4.] There are two distinct components with having different velocities ($\sim 5$ km s$^{-1}$
 and 8 km s$^{-1}$). They appear to converge at the position of the Serpens South and W40 cloud.
 This is consistent with the scenario that the collision between flows and clouds triggered 
active star formation in this region \citep{nakamura14}.
However, there is a possibility that the two components are simply overlapped along the line of sight.

\item[5.] Applying \texttt{Dendrogram}+\texttt{SCIMES} to the $^{13}$CO data cube, 
we identified 61 clouds.

\item[6.] 
The line-width-radius relation of the clouds reasonably 
agrees with those of nearby star-forming regions. However, if we adopt the uniform distance of 415 pc to the region, 
the agreement of the line-width-radius relations becomes worse. We speculate that the representative distance to this area
is $d\sim 260$ pc, or two components with different distances ($\sim$ 260 pc and $\sim $ 415 pc) are overlapped 
along the line of sight.

\item[7.] The virial-parameter-mass relation shows that the clouds identified are close to
virial equilibrium with large dispersion.
This may be the reason why the observed area contains several active star forming regions.
This characteristic is contrast to that of  the southern part of Aquila Rift, where
the virial parameters tend to be larger than unity \citep{kawamura99}.
For a larger assumed distance of $d\sim$ 415 pc, 
the majority of the clouds have virial parameter smaller than unity. In other words, 
the clouds appear too gravitationally-bounded.

\end{itemize}

\acknowledgements
This research made use of two Python packages:
\texttt{Dendrogram}  \citep{rosolowsky08} and 
\texttt{SCIMES} version 0.2.0 \citep{colombo15}.
The former is a Python package to characterize the hierarchical structure
of molecular cloud. 
The latter is  a Python package to find relevant structures into dendrograms 
of molecular gas emission using the spectral clustering approach.
We made use of Astropy version 1.1.2 python package to make several plots \citep{astropy}.
Without these packages, we could not complete the analyses presented in this paper.
We would like to express our special appreciation to all the developers of the packages. 
We also thank to Kazuki Tokuda who helped data reduction of the 1.85-m data.

\clearpage

\begin{deluxetable}{llllllll}
\tabletypesize{\scriptsize}
\tablecolumns{6}
\tablecaption{Observed lines}
\tablewidth{\columnwidth}
\tablehead{\colhead{Molecule} 
& \colhead{Transition} 
& \colhead{Frequency (GHz)$^a$} 
& \colhead{Beam (arcmin)}  
& \colhead{$\Delta V$ (km s$^{-1}$)} 
& \colhead{$\Delta T_{\rm mb}$} (K)}
\startdata
   $^{12}$CO  &  $J=2-1$ & 230.5380000  & 2.7  & 0.079 & 0.64 $\pm$ 0.10  \\
$^{13}$CO  &  $J=2-1$ & 220.3986765  & 2.7  & 0.083 & 0.62 $\pm$ 0.10  \\
C$^{18}$O  &  $J=2-1$ & 219.5603568  & 2.7  & 0.083 & 0.62 $\pm$ 0.10 
\enddata
\tablecomments{The last column is average rms noise levels of the whole area. 
The size of each observation box is $1^\circ \times 1^\circ$, whose
noise level varies from box to box. Therefore, the rms noise level is
measured in each observation box, and is indicated with the standard deviation.}
\label{tab:obs}
\end{deluxetable}

\begin{deluxetable}{llllllllll}
\tabletypesize{\scriptsize}
\tablecolumns{10}
\tablecaption{Clouds identified by the \texttt{SCIMES}}
\tablewidth{\columnwidth}
\tablehead{\colhead{id} & \colhead{Gal. Long.}  & \colhead{Gal. Lat.} & 
\colhead{R.A.} & \colhead{Del.}  & \colhead{$R_{\rm maj}$} & \colhead{$R_{\rm min}$}  
&  \colhead{Position angle} & \colhead{Radius} & \colhead{Velocity} \\
\colhead{}    &  \colhead{(degree)} & \colhead{(degree)} &  \colhead{(J2000)} &  \colhead{(J2000)} & 
\colhead{(arcsec)}  & \colhead{(arcsec)} & \colhead{(degree)} & \colhead{(arcsec)} & \colhead{(km s$^{-1}$)}
}
\startdata
1 & 25.963  &  4.113  & 18h24m01.1s & -04d18m04.1s & 3354.1 & 1422.7 & 89.9 & 2184.4 & 3.6  \\
2 & 31.118  &  5.203  & 18h29m36.6s & +00d45m41.5s & 3621.6 & 1125.6 & 156.9 & 2019.0 & 8.0  \\
3 & 30.567  &  3.202  & 18h35m43.7s & -00d38m28.9s & 3124.3 & 597.1 & 173.6 & 1365.8 & 9.5  \\
4 & 31.822  &  2.936  & 18h38m57.7s & +00d21m07.3s & 1449.3 & 1330.8 & 124.9 & 1388.8 & 8.7  \\
5 & 31.336  &  4.410  & 18h32m49.7s & +00d35m34.5s & 1552.5 & 771.4 & 55.4 & 1094.4 & 10.3 \\
6 & 27.094  &  4.131  & 18h26m02.9s & -03d17m34.7s & 1042.7 & 671.2 & 91.4 & 836.6 & 3.3  \\
7 & 31.576  &  3.958  & 18h34m52.5s & +00d35m59.8s & 562.4 & 325.7 & 157.1 & 428.0 & 6.7  \\
8 & 27.136  &  4.763  & 18h23m53.0s & -02d57m49.8s & 5639.3 & 737.4 & 58.3 & 2039.2 & 8.3  \\
9 & 26.874  &  3.514  & 18h27m50.1s & -03d46m21.2s & 2390.2 & 1551.0 & 159.1 & 1925.4 & 5.4  \\
10 & 27.529  &  5.652  & 18h21m27.0s & -02d12m21.3s & 1576.8 & 762.2 & 173.4 & 1096.3 & 8.0  \\
11 & 25.811  &  3.066  & 18h27m27.7s & -04d55m16.2s & 3181.2 & 1132.9 & 158.2 & 1898.4 & 8.7  \\
12 & 30.420  &  2.482  & 18h38m01.3s & -01d06m04.8s & 1552.7 & 1294.8 & 165.7 & 1417.9 & 8.1  \\
13 & 26.068  &  4.048  & 18h24m26.8s & -04d14m18.0s & 1525.3 & 973.9 & 163.2 & 1218.8 & 7.9  \\
14 & 32.384  &  2.794  & 18h40m29.7s & +00d47m11.8s & 871.1 & 465.8 & 162.8 & 637.0 & 10.0 \\
15 & 32.184  &  2.332  & 18h41m46.4s & +00d23m54.0s & 1068.8 & 506.7 & 165.3 & 735.9 & 9.6  \\
16 & 28.949  &  3.766  & 18h30m45.5s & -01d49m06.7s & 2659.8 & 2070.8 & 73.1 & 2346.9 & 7.1  \\
17 & 32.213  &  3.347  & 18h38m12.6s & +00d53m11.1s & 781.8 & 334.8 & 179.6 & 511.7 & 7.9  \\
18 & 32.760  &  2.718  & 18h41m27.0s & +01d05m11.1s & 987.5 & 732.3 & 176.6 & 850.4 & 11.7 \\
19 & 28.849  &  5.550  & 18h24m14.1s & -01d05m13.9s & 1225.3 & 673.4 & 115.6 & 908.4 & 8.5  \\
20 & 30.417  &  4.672  & 18h30m13.5s & -00d06m09.7s & 725.0 & 295.8 & 155.0 & 463.1 & 7.8  \\
21 & 31.291  &  3.482  & 18h36m03.2s & +00d07m46.1s & 1052.2 & 600.6 & 105.0 & 794.9 & 6.4  \\
22 & 25.465  &  1.944  & 18h30m49.0s & -05d44m47.8s & 1704.6 & 508.9 & 165.1 & 931.4 & 10.0 \\
23 & 27.484  &  1.851  & 18h34m52.7s & -03d59m52.9s & 1727.6 & 1148.3 & 155.0 & 1408.5 & 8.8 \\
24 & 25.368  &  4.271  & 18h22m21.1s & -04d45m12.6s & 672.3 & 613.4 & 84.2 & 642.2 & 6.6  \\
25 & 32.894  &  2.189  & 18h43m34.5s & +00d57m53.5s & 457.4 & 344.4 & 161.3 & 396.9 & 11.8 \\
26 & 25.125  &  3.864  & 18h23m20.9s & -05d09m26.1s & 694.5 & 380.0 & 141.3 & 513.7 & 7.2  \\
27 & 26.678  &  2.616  & 18h30m40.1s & -04d21m39.0s & 1609.7 & 771.2 & 94.7 & 1114.2 & 5.6  \\
28 & 30.888  &  4.145  & 18h32m57.3s & +00d04m28.2s & 687.7 & 569.2 & 104.1 & 625.6 & 6.6  \\
29 & 29.677  &  5.169  & 18h27m06.3s & -00d31m46.1s & 1502.9 & 724.6 & 53.2 & 1043.6 & 8.8  \\
30 & 25.092  &  5.561  & 18h17m15.4s & -04d23m42.0s & 507.4 & 382.4 & 117.9 & 440.5 & 3.0  \\
31 & 32.822  &  3.532  & 18h38m39.7s & +01d30m42.6s & 673.5 & 549.8 & 136.9 & 608.5 & 9.3  \\
32 & 28.617  &  5.234  & 18h24m56.0s & -01d26m16.6s & 2363.8 & 683.9 & 165.5 & 1271.5 & 5.4  \\
33 & 27.167  &  5.120  & 18h22m40.3s & -02d46m15.7s & 1309.4 & 920.8 & 167.0 & 1098.0 & 3.9  \\
34 & 32.359  &  2.507  & 18h41m28.2s & +00d37m59.9s & 436.0 & 174.1 & 159.6 & 275.5 & 8.8  \\
35 & 32.715  &  2.742  & 18h41m16.8s & +01d03m24.7s & 740.4 & 448.8 & 110.5 & 576.5 & 10.0 \\
36 & 30.030  &  1.919  & 18h39m18.5s & -01d42m20.9s & 1561.6 & 797.0 & 161.0 & 1115.6 & 7.4  \\
37 & 27.910  &  3.273  & 18h30m36.2s & -02d57m59.2s & 1553.8 & 491.9 & 83.0 & 874.3 & 6.3  \\
38 & 30.371  &  4.139  & 18h32m02.2s & -00d23m13.0s & 379.0 & 354.9 & 156.0 & 366.7 & 9.3  \\
39 & 31.161  &  5.882  & 18h27m16.3s & +01d06m35.3s & 842.3 & 370.9 & 173.8 & 558.9 & 7.8  \\
40 & 32.700  &  2.295  & 18h42m50.7s & +00d50m23.8s & 1035.9 & 293.5 & 155.2 & 551.4 & 10.1 \\
41 & 25.199  &  5.069  & 18h19m12.3s & -04d31m52.3s & 1556.9 & 938.7 & 55.3 & 1208.9 & 6.2  \\
42 & 25.861  &  3.192  & 18h27m06.2s & -04d49m06.0s & 849.9 & 355.6 & 164.4 & 549.8 & 2.5  \\
43 & 30.493  &  4.323  & 18h31m36.2s & -00d11m40.1s & 1317.9 & 511.5 & 168.7 & 821.0 & 8.0  \\
44 & 25.455  &  3.817  & 18h24m07.8s & -04d53m14.2s & 1175.3 & 649.9 & 169.4 & 874.0 & 6.9  \\
45 & 25.139  &  5.629  & 18h17m06.3s & -04d19m22.3s & 735.9 & 437.5 & 96.8 & 567.4 & 4.1  \\
46 & 25.689  &  4.375  & 18h22m34.7s & -04d25m17.4s & 696.2 & 286.7 & 150.5 & 446.8 & 4.8  \\
47 & 32.226  &  2.389  & 18h41m38.8s & +00d27m42.3s & 588.8 & 429.3 & 136.8 & 502.7 & 5.4  \\
48 & 32.035  &  3.662  & 18h36m46.0s & +00d52m20.1s & 433.2 & 209.2 & 161.3 & 301.0 & 6.1  \\
49 & 31.241  &  4.303  & 18h33m02.3s & +00d27m35.1s & 440.1 & 279.9 & 141.7 & 351.0 & 6.2  \\
50 & 32.311  &  3.911  & 18h36m22.9s & +01d13m51.3s & 412.1 & 187.4 & 173.8 & 277.9 & 6.4  \\
51 & 25.350  &  3.620  & 18h24m37.9s & -05d04m18.7s & 512.5 & 275.8 & 108.8 & 376.0 & 7.8  \\
52 & 25.741  &  1.885  & 18h31m32.2s & -05d31m42.5s & 408.3 & 312.1 & 60.3 & 357.0 & 9.6  \\
53 & 26.925  &  3.050  & 18h29m34.9s & -03d56m29.7s & 630.8 & 470.6 & 164.2 & 544.8 & 9.7  \\
54 & 30.558  &  4.922  & 18h29m35.4s & +00d08m15.6s & 628.2 & 355.6 & 147.8 & 472.6 & 9.9  \\
55 & 26.457  &  3.181  & 18h28m14.8s & -04d17m44.6s & 797.4 & 328.5 & 157.4 & 511.8 & 11.7 \\
56 & 30.134  &  4.615  & 18h29m54.6s & -00d22m45.5s & 524.2 & 290.3 & 142.6 & 390.1 & 11.2 \\
57 & 28.148  &  1.093  & 18h38m48.1s & -03d45m21.8s & 598.2 & 277.3 & 168.3 & 407.3 & 11.8 \\
58 & 27.603  &  1.875  & 18h35m00.6s & -03d52m53.2s & 533.4 & 329.1 & 163.9 & 419.0 & 11.7 \\
59 & 32.732  &  2.052  & 18h43m46.1s & +00d45m26.9s & 345.9 & 185.5 & 172.5 & 253.3 & 13.1 \\
60 & 32.092  &  3.376  & 18h37m53.2s & +00d47m31.2s & 320.4 & 247.7 & 124.8 & 281.8 & 13.7 \\
61 & 32.667  &  1.181  & 18h46m45.0s & +00d18m11.7s & 839.8 & 256.9 & 131.8 & 464.5 & 20.2 
\enddata
\label{tab:scime1}
\end{deluxetable}

\begin{deluxetable}{lllll}
\tabletypesize{\scriptsize}
\tablecolumns{5}
\tablecaption{Physical Properties of Clouds identified by the \texttt{SCIMES}}
\tablewidth{\columnwidth}
\tablehead{\colhead{id} & \colhead{Radius} & \colhead{Mass} & \colhead{$\Delta V_{\rm FWHM}$} & 
\colhead{Virial Parameter} \\
\colhead{} & \colhead{(pc)} & \colhead{($M_\odot$) } & \colhead{(km s$^{-1}$)} & \colhead{} 
}
\startdata
1 & 2.8  & 552.5 & 1.51  & 1.4  \\
2 & 2.5  & 3875.1 & 1.78  & 0.3  \\
3 & 1.7  & 584.2 & 2.07  & 1.6  \\
4 & 1.8  & 1722.2 & 2.67  & 0.9  \\
5 & 1.4  & 277.8 & 1.13  & 0.8  \\
6 & 1.1  & 74.6 & 0.91  & 1.5  \\
7 & 0.5  & 31.2 & 0.77  & 1.3  \\
8 & 2.6  & 1005.1 & 2.01  & 1.3  \\
9 & 2.4  & 1984.3 & 3.53  & 1.9  \\
10 & 1.4  & 306.7 & 1.31  & 1.0  \\
11 & 2.4  & 464.4 & 1.46  & 1.4  \\
12 & 1.8  & 1055.2 & 2.27  & 1.1  \\
13 & 1.5  & 308.3 & 1.51  & 1.4  \\
14 & 0.8  & 56.4 & 0.62  & 0.7  \\
15 & 0.9  & 85.7 & 1.16  & 1.8  \\
16 & 3.0  & 6009.1 & 2.75  & 0.5  \\
17 & 0.6  & 46.8 & 0.66  & 0.8  \\
18 & 1.1  & 216.8 & 1.57  & 1.5  \\
19 & 1.1  & 72.9 & 0.95  & 1.8  \\
20 & 0.6  & 7.4  & 0.30  & 0.9  \\
21 & 1.0  & 172.9 & 1.04  & 0.8  \\
22 & 1.2  & 106.7 & 0.94  & 1.2  \\
23 & 1.8  & 470.8 & 3.03  & 4.3  \\
24 & 0.8  & 40.1 & 0.53  & 0.7  \\
25 & 0.5  & 42.3 & 0.94  & 1.3  \\
26 & 0.6  & 21.7 & 0.86  & 2.8  \\
27 & 1.4  & 470.1 & 1.40  & 0.7  \\
28 & 0.8  & 84.9 & 1.47  & 2.5  \\
29 & 1.3  & 143.9 & 1.23  & 1.7  \\
30 & 0.6  & 14.5 & 0.45  & 1.0  \\
31 & 0.8  & 49.6 & 0.89  & 1.5  \\
32 & 1.6  & 121.6 & 1.22  & 2.5  \\
33 & 1.4  & 73.6 & 0.82  & 1.6  \\
34 & 0.3  & 9.8  & 0.48  & 1.0  \\
35 & 0.7  & 25.4 & 0.45  & 0.7  \\
36 & 1.4  & 372.7 & 1.73  & 1.4  \\
37 & 1.1  & 49.2 & 0.43  & 0.5  \\
38 & 0.5  & 22.0 & 1.03  & 2.8  \\
39 & 0.7  & 34.0 & 0.45  & 0.5  \\
40 & 0.7  & 10.1 & 0.33  & 1.0  \\
41 & 1.5  & 177.5 & 1.11  & 1.3  \\
42 & 0.7  & 39.7 & 0.81  & 1.5  \\
43 & 1.0  & 72.8 & 0.77  & 1.1  \\
44 & 1.1  & 42.6 & 0.50  & 0.8  \\
45 & 0.7  & 31.7 & 0.73  & 1.5  \\
46 & 0.6  & 12.8 & 0.64  & 2.3  \\
47 & 0.6  & 19.1 & 0.63  & 1.7  \\
48 & 0.4  & 11.4 & 0.83  & 2.9  \\
49 & 0.4  & 6.7  & 0.42  & 1.5  \\
50 & 0.4  & 12.3 & 0.47  & 0.8  \\
51 & 0.5  & 5.1  & 0.23  & 0.6  \\
52 & 0.4  & 10.6 & 0.49  & 1.3  \\
53 & 0.7  & 15.9 & 0.45  & 1.1  \\
54 & 0.6  & 14.5 & 0.31  & 0.5  \\
55 & 0.6  & 98.7 & 1.55  & 2.0  \\
56 & 0.5  & 14.4 & 0.76  & 2.4  \\
57 & 0.5  & 20.8 & 1.22  & 4.6  \\
58 & 0.5  & 6.7  & 0.22  & 0.5  \\
59 & 0.3  & 17.5 & 0.79  & 1.4  \\
60 & 0.4  & 14.9 & 0.96  & 2.7  \\
61 & 0.6  & 154.1 & 1.17  & 0.6  
\enddata
\label{tab:scime2}
\end{deluxetable}

\clearpage

\clearpage

\appendix

\section{The noise levels of the data}

The rms noise levels of the final $^{12}$CO, $^{13}$CO, and C$^{18}$O maps 
varies from box to box, and were summarized in Tables A1, A2, and  A3, respectively.
The average noise levels are summarized in the last column of Table 1.

\begin{deluxetable}{lllllllll}
\tabletypesize{\scriptsize}
\tablecolumns{9}
\tablecaption{Rms noise levels of the $^{12}$CO map}
\tablewidth{\columnwidth}
\tablehead{\colhead{} & \colhead{$l=33^\circ-32^\circ$}& \colhead{$l=32^\circ-31^\circ$} & \colhead{$l=31^\circ-30^\circ$} & 
\colhead{$l=30^\circ-29^\circ$} & \colhead{$l=29^\circ-28^\circ$} & \colhead{$l=28^\circ-27^\circ$} & \colhead{$l=27^\circ-26^\circ$} & \colhead{$l=26^\circ-25^\circ$} }
\startdata
$b=6^\circ-5^\circ$ & 0.75 & 0.66  & 0.76  & 0.53 & 0.58  & 0.57 & 0.50 & 0.69 \\
$b=5^\circ-4^\circ$ &  0.65 & 0.65 & 0.80 & 0.43 & 0.45  & 0.44 & 0.49 & 0.71 \\
$b=4^\circ-3^\circ$ &  0.87 & 0.69  & 0.57 & 0.35 & 0.28  & 0.44 & 0.63 & 0.80 \\
$b=3^\circ-2^\circ$ &  0.83 & 0.53  & 0.51 & 0.41 & 0.45  & 0.44 & 0.52 & 0.57 \\
$b=2^\circ-1^\circ$ &  0.54 & 0.56  & 0.55 & 0.70 & 0.68  & 0.61 & 0.53 & 0.63 
\enddata
\tablecomments{The rms noise levels are given in the brightness temperature scale (K).
The velocity resolution is 0.079 km s$^{-1}$.}
\label{tab:noise12co}
\end{deluxetable}
 
\begin{deluxetable}{lllllllll}
\tabletypesize{\scriptsize}
\tablecolumns{9}
\tablecaption{Rms noise levels of the $^{13}$CO map}
\tablewidth{\columnwidth}
\tablehead{\colhead{} & \colhead{$l=33^\circ-32^\circ$}& \colhead{$l=32^\circ-31^\circ$} & \colhead{$l=31^\circ-30^\circ$} & 
\colhead{$l=30^\circ-29^\circ$} & \colhead{$l=29^\circ-28^\circ$} & \colhead{$l=28^\circ-27^\circ$} & \colhead{$l=27^\circ-26^\circ$} & \colhead{$l=26^\circ-25^\circ$} }
\startdata
$b=6^\circ-5^\circ$ & 0.72 & 0.64  & 0.73  & 0.52 & 0.56  & 0.55 & 0.49 & 0.67 \\
$b=5^\circ-4^\circ$ &  0.63 & 0.64 & 0.77 & 0.41 & 0.43  & 0.42 & 0.48 & 0.68 \\
$b=4^\circ-3^\circ$ &  0.85 & 0.68  & 0.56 & 0.35 & 0.28  & 0.43 & 0.61 & 0.77 \\
$b=3^\circ-2^\circ$ &  0.81 & 0.52  & 0.50 & 0.39 & 0.42  & 0.41 & 0.56 & 0.56 \\
$b=2^\circ-1^\circ$ &  0.53 & 0.54  & 0.53 & 0.68 & 0.66  & 0.59 & 0.51 & 0.61 \\
\enddata
\tablecomments{The rms noise levels are given in the brightness temperature scale (K).
The velocity resolution is 0.083 km s$^{-1}$.}
\label{tab:noise13co}
\end{deluxetable}

\begin{deluxetable}{lllllllll}
\tabletypesize{\scriptsize}
\tablecolumns{9}
\tablecaption{Rms noise levels of the C$^{18}$O map}
\tablewidth{\columnwidth}
\tablehead{\colhead{} & \colhead{$l=33^\circ-32^\circ$}& \colhead{$l=32^\circ-31^\circ$} & \colhead{$l=31^\circ-30^\circ$} & 
\colhead{$l=30^\circ-29^\circ$} & \colhead{$l=29^\circ-28^\circ$} & \colhead{$l=28^\circ-27^\circ$} & \colhead{$l=27^\circ-26^\circ$} & \colhead{$l=26^\circ-25^\circ$} }
\startdata
$b=6^\circ-5^\circ$ & 0.73 & 0.64  & 0.73  & 0.52 & 0.56  & 0.55 & 0.49 & 0.67 \\
$b=5^\circ-4^\circ$ &  0.64 & 0.64 & 0.77 & 0.41 & 0.43  & 0.41 & 0.48 & 0.68 \\
$b=4^\circ-3^\circ$ &  0.84 & 0.68  & 0.56 & 0.34 & 0.28  & 0.43 & 0.60 & 0.78 \\
$b=3^\circ-2^\circ$ &  0.81 & 0.52  & 0.50 & 0.39 & 0.42  & 0.41 & 0.55 & 0.55 \\
$b=2^\circ-1^\circ$ &  0.53 & 0.54  & 0.53 & 0.68 & 0.65  & 0.58 & 0.50 & 0.61 
\enddata
\tablecomments{The rms noise levels are given in the brightness temperature scale (K).
The velocity resolution is 0.083 km s$^{-1}$.}
\label{tab:noisec18o}
\end{deluxetable}

\end{document}